\documentclass[useAMS,usenatbib]{mn2e}
\usepackage{times} 
\usepackage{graphicx}
\usepackage{rotating}
\usepackage{epsfig}
\usepackage{amsmath}
\def\PfB2{{P$_f$B2}}
\def\PlB3{{P$_l$B3}}
\def\PlhA3{{P$_l$A3}}
\def\PlA4{{P$_l$A4}}
\def\td{{\em Triax+Disk}}
\def\pe{{\em Prolt+Ellip}}
\def\pbbh{{\em Prolt+hardpt}}
\def\tb{{\em Triax+Bulg}}
\def\tbbh{{\em Triax+hardpt}}

\def\msun{\ifmmode{{\mathrm M}_{\odot}}\else{M$_{\odot}$}\fi} 
\def\kms{\ifmmode{{\mathrm{km \, s^{-1}}}}\else{${\mathrm{km \, s^{-1}}}$}\fi} 

\title{The Orbital Evolution Induced by Baryonic Condensation in Triaxial Halos}
\author[Valluri et al.]{Monica Valluri$^{1}$\footnote{E-mail:mvalluri@umich.edu (MV); ; vpdebattista@uclan.ac.uk (VPD)},
Victor P. Debattista$^{2}$, Thomas Quinn$^{3}$, Ben Moore$^{4}$\\
$^{1}$ Department of Astronomy, University of Michigan, Ann Arbor, MI 48109, USA\\
$^{2}$ RCUK Fellow; Jeremiah Horrocks Institute, University of Central Lancashire,  Preston, PR1 2HE, UK\\
$^{3}$ Astronomy Department, University of Washington, Box 351580, Seattle, WA 98195-1580, USA\\
$^{4}$ Department of Theoretical Physics, University of Z\"urich, Winterthurerstrasse 190, CH-8057, Z\"urich, Switzerland\\}
\begin{document}

\date{Accepted - Dec, 11, 2009}

\pagerange{\pageref{firstpage}--\pageref{lastpage}} \pubyear{2009}

\maketitle

\label{firstpage}

\begin{abstract} 

Using spectral methods, we analyse the orbital structure of
prolate/triaxial dark matter (DM) halos in $N$-body simulations in an
effort to understand the physical processes that drive the evolution
of shapes of dark matter halos and elliptical galaxies in which
central masses are grown. A longstanding issue is whether the change
in the shapes of DM halos is the result of chaotic scattering of the
major family of box orbits that serves as the back-bone of a triaxial
system, or whether they change shape adiabatically in response to the evolving
galactic potential.  We use the characteristic orbital frequencies to
classify orbits into major orbital families, to quantify orbital
shapes, and to identify resonant orbits and chaotic orbits. The use 
of a frequency-based method for distinguishing between regular and 
chaotic $N$-body orbits overcomes the limitations of Lyapunov 
exponents which are sensitive to numerical discreteness effects. We show
that regardless of the distribution of the baryonic component, the
shape of a DM halo changes primarily due to changes in the shapes of
individual orbits within a given family.  Orbits with small
pericentric radii are more likely to change both their orbital type
and shape than orbits with large pericentric radii. Whether the
evolution is regular (and reversible) or chaotic (and irreversible),
depends primarily on the radial distribution of the baryonic
component.   The growth of an extended baryonic component of 
any shape results in aregular and reversible change in orbital
 populations and shapes, features that are not expected for 
 chaotic evolution. In contrast the growth of a massive and compact 
 central component results in chaotic scattering of a significant fraction 
 of both box and long-axis tube orbits, even those with pericenter 
 distances  much larger than the size of the central component. 
 Frequency maps show that the growth of a disk causes a significant 
 fraction of halo particles to become trapped by major global orbital 
resonances. We find that despite the fact that shape of a DM halo is 
always quite oblate following the growth of a central baryonic 
component, a significant fraction of its orbit population has the 
characteristics of its triaxial or prolate progenitor.

\end{abstract}


\section{Introduction}
\label{sec:intro}

The condensation of baryons to the centres of dark matter halos is
known to make them more spherical or axisymmetric \citep[][hereafter
D08]{, deb_etal_08}.  D08 found that the halo shape
changes by $\Delta(b/a) \ga 0.2$ out to at least half the virial
radius.  This shape change reconciles the strongly prolate-triaxial
shapes found in collisionless $N$-body simulations of the hierarchal
growth of halos \citep{bbks_86, bar_efs_87, frenk_etal_88, dub_car_91,
jin_sut_02, bai_ste_05, all_etal_06} with observations, which
generally find much rounder halos \citep{sch_etal_83, sac_spa_90,
fra_dez_92, hui_van_92, kui_tre_94, fra_etal_94, buo_can_94,
bar_etal_95, kochan_95, olling_95, olling_96, sch_etal_97,
koo_etal_98, oll_mer_00, and_etal_01, buo_etal_02, ogu_etal_03,
bar_sel_03, debatt_03, iod_etal_03, die_sta_07, ban_jog_08}.

What is the physical mechanism driving shape change? Options suggested
in the literature include two possibilities.  The first is that the
presence of a central mass concentration scatters box orbits that
serve as the backbone of a triaxial potential, rendering them chaotic
\citep{ger_bin_85,merritt_valluri_96}.  Chaotic orbits in a stationary
potential do not conserve any integrals of motion other than the
energy $E$ and consequently are free to uniformly fill their allowed
equipotential surface. Since the potential is, in general, rounder
than the density distribution chaotic diffusion results in evolution
to a more oblate or even a spherical shape \citep{mer_qui_98,
kalapotharkos_08}.  The second possibility is that the change of the
central potential occurs because the growth of the baryonic component
causes orbits of collisionless particles in the halo to respond by
changing their shapes in a regular (and therefore reversible) manner
\citep{hol_etal_02}.

Time dependence in a potential is also believed to result in chaotic
mixing \citep{terzic_kandrup_04, kandrup_novotny_04} and has been
invoked as the mechanism that drives violent relaxation. However, a
more recent analysis of mixing during a major merger showed that the
rate and degree of mixing in energy and angular momentum are not
consistent with chaotic mixing, but rather that particles retain
strong memory of their initial energies and angular momenta even in
strongly time dependent potentials \citep{valluri_etal_07}.

One of the principal features of chaotic evolution is
irreversibility. This irreversibility arises from two properties of
chaotic orbits. First, chaotic orbits are exponentially sensitive to
small changes in initial conditions even in a collisionless
system. Second, chaotic systems display the property of {\it chaotic
mixing} \citep{lichtenberg_lieberman_92}. Using this principle of
irreversibility, D08 argued that if chaotic evolution is the primary
driver of shape change, then if, subsequently, the central mass
concentration is artificially ``evaporated'', the system would not be
able to revert to its original triaxial distribution.  D08 showed that
growing baryonic components inside prolate/triaxial halos led to a
large change in the shape of the halo.  Despite these large changes,
by artificially evaporating the baryons, they showed that the
underlying halo phase space distribution is not grossly altered unless
the baryonic component is too massive or centrally concentrated, or
transfers significant angular momentum to the halo.  This led them to
argue that chaotic evolution alone cannot explain the shape change
since such a process is irreversible.  They speculated that at most
only slowly diffusive chaos occurred in their simulations.  Using test
particle orbit integrations they also showed that box orbits largely
become deformed, possibly changing into tube orbits, during disk
growth, but do not become strongly chaotic.

D08 employed irreversibility as a convenient proxy for the presence of
chaos.  In this paper we undertake an orbital analysis of some of the
models studied by D08 to better understand the mechanism that drives
shape change.  Our goal is to understand whether chaotic orbits are an
important driver of shape change and if so under what conditions they
are important. We also wish to understand how the orbital populations
in halos change when a centrally concentrated baryonic component grows
inside a triaxial dark matter halo. Finally we would like to
understand under what circumstances orbits change their
classification.

This paper is organised as follows. In \S~\ref{sec:simulations} we
describe the simulations used in this paper and briefly describe three
models from D08 as well as two additional simulations. In
\S~\ref{sec:frequency} we describe the principal technique: Numerical
Analysis of Fundamental Frequencies (NAFF) that we use to obtain
frequency spectra and fundamental frequencies and describe how these
frequencies are used to characterise orbits. In \S~\ref{sec:results}
we describe the results of our analysis of five different simulations.
In \S~\ref{sec:summary} we summarise our results and discuss their
implications.


\section{Numerical Simulations}
\label{sec:simulations}

\begin{table*}
\begin{centering}
\begin{tabular}{cccccccccc}\hline 
\multicolumn{1}{c}{Run Number} &
\multicolumn{1}{c}{Run }&
\multicolumn{1}{c}{Halo} &
\multicolumn{1}{c}{$r_{200}$} &
\multicolumn{1}{c}{$M_{200}$} &
\multicolumn{1}{c}{$M_b$} &
\multicolumn{1}{c}{$f_b$}&
\multicolumn{1}{c}{$R_b$} &
\multicolumn{1}{c}{$t_g$} &
\multicolumn{1}{c}{$t_e$} \\ 

(from D08) & Description  &  & [kpc] & [$10^{12} \msun$] & [$10^{11} \msun$] &  & [kpc] & [Gyr] & [Gyr] \\ \hline
SA1         & {\em Triax+Disk}  & A & 215 &   4.5 & 1.75  & 0.039  & 3.0 &  5  & 2.5 \\ 
P$_l$A3 & {\em Triax+Bulg}      & A & 215 &   4.5 & 1.75  & 0.039  & 1.0 &  5  & 2.5 \\
P$_f$B2 & {\em Prolt+Ellip}      & B & 106 & 0.65 & 0.7   & 0.108   & 3.0 &10  & 4 \\ 
P$_l$A4 &  {\em Triax+hardpt}     & A & 215 &   4.5 & 1.75  & 0.039  & 0.1 &  5  & 2.5  \\
P$_l$B3 &  {\em Prolt+hardpt}   & B & 106 & 0.65 & 0.35 & 0.054  & 0.1 &   5 & 5 \\ 
\hline
\end{tabular}
\caption{The simulations in this paper.  $M_b$ is the mass in baryons and $f_b$ is 
the baryonic mass fraction. For the particle simulations (P$_f$B2, P$_l$B3, P$_l$A3, P$_l$A4), 
$R_{\rm b}$ refers to the softening of
the spherical baryonic distribution particle(s). For simulation SA1,
$R_{\rm b}$ refers to the scale length of the baryonic disk.}
\label{tab:simulations}
\end{centering}
\end{table*}

We formed prolate/triaxial halos via mergers of systems, as described
in \citet{moo_etal_04}.  The initially spherical NFW \citep{nfw} halos
were generated from a distribution function using the method described
in \citet{kmm04} with each halo composed of two mass species arranged
on shells.  The outer shell has more massive particles than the inner
one, similar to the method described by \citet{zemp_etal_08}, which
allows for higher mass resolution at small radii.  Our model halo A
was generated by the head-on merger of two prolate halos, themselves
the product of a binary merger of spherical systems.  The first merger
placed the concentration $c=10$ halos 800 kpc apart approaching each
other at 50 \kms, while the second merger starts with the remnant at
rest, 400 kpc from an identical copy.  The resulting halo is highly
prolate with a mild triaxiality.  Halo model B was produced by the
merger of two spherical halos starting at rest, 800 kpc apart and is
prolate, with $\left<b/a\right> = \left<c/a\right> \simeq 0.58$.  Halo
A has $\left<b/a\right> \simeq 0.45$ and $\left<c/a\right> \simeq
0.35$ while halo B has $\left<b/a\right> = \left<c/a\right> \simeq
0.58$ (see Figure 3 of D08 for more details).  Both halos A and B
consist of $4\times 10^6$ particles.  The outer particles are $\sim
18$ times more massive in halo A and $\sim 5$ times more massive in
halo B.  A large part of the segregation by particle mass persists
after the mergers and the small radius regions are dominated by low
mass particles \citep[cf.][]{dehnen_05}.  We used a softening
parameter $\epsilon = 0.1$ kpc for all halo particles.  The radius,
$r_{200}$, at which the halo density is 200 times the mean density of
the Universe and the total mass within this radius, $M_{200}$, are
given in Table 1.

Once we produced the prolate/triaxial halos, we inserted a baryonic
component, either a disk of particles that remains rigid throughout
the experiments or softened point particles.  The parameters that
describe the distribution of the baryonic components are given in
Table1. In four of the models (\PlhA3, \PlA4, \PfB2 and \PlB3) the
baryonic component is simply a softened point mass with softening
scale length given by $R_b$.  In model SA1, the density distribution
of the disk was exponential with scale length of the baryonic
component $R_{\rm b}$ and Gaussian scale-height $z_{\rm b}/R_{\rm b} =
0.05$.  The disk was placed with its symmetry axis along the triaxial
halo's short axis in model SA1 (additional orientations of the disk
relative to the principal axes were also simulated but their
discussion is deferred to a future paper).  Initially, the disk has
negligible mass, but it grows adiabatically and linearly with time to
a mass $M_b$ during a time $t_g$.  After this time, we slowly
evaporated it during a time $t_e$.  We stress that this evaporation is
a numerical convenience for testing the effect of chaos on the system,
and should not be mistaken for a physical evolution.  The disk is
composed of $300K$ equal-mass particles each with a softening
$\epsilon = 100$ pc.  From $t=0$ to $t_g+t_e$ the halo particles are
free to move and achieve equilibrium with the baryons as their mass
changes, but all disk particles are frozen in place.  The masses of
models with single softened particles are also grown in the same way;
these are models \PfB2, \PlB3 from D08, and \PlhA3 and \PlA4 which are
new to this paper.  D08's naming convention for these experiments used
``P'' subscripted by ``f'' for particles frozen in place and by ``l''
for live particles free to move.  

Three different baryonic components are grown in the triaxial halo A:
in model SA1 the baryons are in the form of a disk grown perpendicular
to the short axis and the model is referred to as {\em Triax+Disk}; in
model P$_l$A3 the baryonic component is a softened central point mass
resembling a bulge and the model is referred to as {\em Triax+Bulg};
finally in model P$_l$A4 the baryonic component is a hard central
point mass with a softening of 0.1~kpc and the model is referred to as
{\em Triax+hardpt}. Two different baryonic components are grown in the
prolate halo B: in model P$_f$B2 the baryonic component loosely
resembles an elliptical galaxy so this run is referred to as {\em
Prolt+Ellip}; in model P$_l$B3 the baryonic component is a hard
central point mass with a softening of 0.1~kpc and is referred to as
{\em Prolt+hardpt}. For model \PfB2, D08 showed that
there is no significant difference in the evolution if the central
particle is live instead of frozen, all other things being equal.
 P$_l$A3 was constructed specifically for this
paper in order to have a triaxial halo model with a moderately soft
spherical baryonic distribution which can be contrasted with the
prolate halo model P$_f$B2, while P$_l$A4 is a triaxial halo model
which can be contrasted with the prolate halo model P$_l$B3.

For each model there were 5 phases in evolution. The initial triaxial
or prolate halo without the baryonic component is referred to as {\it
phase a}. There is then a phase (of duration $t_g$) during which the
halo's shape is evolving as the baryonic component is grown
adiabatically. We do not study orbits in this phase since the
potential is evolving with time.  When the baryonic component has
finished growing to full strength, and the halo has settled to a new
equilibrium the model is referred to as being in {\it phase b}.  The
baryonic component is ``adiabatically evaporated" over a timescale of
duration $t_e$ listed in Table~\ref{tab:simulations}. Again we do not
study orbits during this period when the potential is evolving with
time. After the baryonic component has been adiabatically evaporated
completely and the halo has returned to an equilibrium configuration
the halo is referred to as being in {\it phase c}. We only study the
orbits in the halo during the three phases when the halo is in
equilibrium and is stationary (not evolving with time).

The growth of the baryonic component induces several changes in the
distributions of the DM halo particles: first is an increase in
central density relative to the original NFW halo due to increase in
the depth of the central potential (an effect commonly referred to as
``baryonic compression"); second, the halos become more oblate
especially within 0.3$r_{vir}$.  The details of the changes in the
density and velocity distributions of DM particles differ slightly
depending on the nature of the baryonic component (D08).

All the simulations in this paper, which are listed in Table
\ref{tab:simulations}, were evolved with {\sc pkdgrav} an efficient,
multi-stepping, parallel tree code \citep{stadel_phd}.  We used cell
opening angle for the tree code of $\theta = 0.7$
throughout\footnote{Opening angle $\theta$ is used in tree codes to
determine how long-range forces from particles acting at a point are
accumulated \citep{bar_hut_86}. }. Additional details of the
simulations can be found in D08.

\subsection{Computing Orbits}

In each of the halos studied we selected a subsample of between
1000-6000 particles and followed their orbits in each of the three
stationary phases of the evolution described in the previous section.
The particles were randomly chosen in the halos at $t=0$ such that
they were inside a fixed outer radius (either 100 or 200~kpc).  Since
the particles were selected at random from the distribution function,
they have the same overall distribution as the entire distribution
function within the outer radius selected. We integrated the motion of
each a test particle while holding all
the other particles fixed in place.  We used a fixed timestep of 0.1
Myr and integrated for 50 Gyr, storing the phase space coordinates of
each test particle every 1 Myr.  We used such long integration times
to ensure we are able to obtain accurate measurements of frequencies
(as described in the next section).  We carried out this operation for
the same subset of particles at {\it phases a}, {\it b} and {\it c}.
In model SA1 ({\em Triax+Disk}) we integrated the orbits of 6000
particles which in {\it phase a} were within $r=200$~kpc.  In model
\PlhA3 ({\em Triax+Bulg}) and \PlA4 ({\em Triax+hardpt}) we considered
a subsample of 5000 particles starting within $r=100$~kpc.  In models
\PfB2 and \PlB3 we considered orbits of 1000 particles within
$r=200$~kpc.  We integrated their orbits as above but we used a
smaller timestep $\delta t = 10^4 $ years in the case of \PlA4 and
\PlB3, which had harder central point masses.  The orbit code computes
forces in a frozen potential using an integration scheme that uses forces calculated from  the
{\sc PKDGRAV} tree; we used the orbit integration parameters identical
to those used for the evolution of the self-consistent models.


\section{Frequency Analysis}
\label{sec:frequency}

In a 3-dimensional galactic potential that is close to integrable, all
orbits are quasi-periodic.  If an orbit is quasi-periodic (or
regular), then any of its coordinates can be described explicitly as a
series,
\begin{eqnarray}
x(t) = \sum_{k=1}^{\infty} A_{k} e^{i\omega_kt},
\end{eqnarray}
where the $\omega_k$'s are the oscillation frequencies and the $A_k$'s
are the corresponding amplitudes. In a three dimensional potential,
each $\omega_k$ can be written as an integer linear combination of
three fundamental frequencies $\omega_1, \omega_2, \omega_3$ (one for
each degree of freedom).  If each component of the motion of a
particle in the system (e.g. $x(t)$) is followed for several ($\sim
100$) dynamical times, a Fourier transform of the trajectory yields a
spectrum with discrete peaks. The locations of the peaks in the
spectrum correspond to the frequencies $\omega_k$ and their amplitudes
$A_k$ can be used to compute the linearly independent fundamental
frequencies \citep{boozer_82, kuo-petravic_etal_83, BT08}.

\citet{binney_spergel_82, binney_spergel_84} applied this method to
galactic potentials and obtained the frequency spectra using a least
squares technique to measure the frequencies $\omega_k$.
\citet{laskar_90, laskar_93} developed a significantly improved
numerical technique (Numerical Analysis of Fundamental Frequencies,
hereafter NAFF) to decompose a complex time series of the phase space
trajectory of an orbit of the form $x(t) + i v_x(t)$, (where $v_x$ is
the velocity along the $x$ coordinate). \citet{valluri_merritt_98}
developed their own implementation of this algorithm that uses integer
programming to obtain the fundamental frequencies from the frequency
spectrum. In this paper we use this latter implementation of the NAFF
method.

We refer readers to the above papers and to Section 3.7 of \citet{BT08}
for a detailed discussion of the main idea behind the recovery of
fundamental frequencies. For completeness we provide a brief summary
here. The NAFF algorithm for frequency analysis allows one to quickly
and accurately compute the fundamental frequencies that characterise
the quasi-periodic motion of regular orbits. The entire phase space at
a given energy can then be represented by a frequency map which is a
plot of ratios of the fundamental frequencies of motion. A frequency
map is one of the easiest ways to identify families of orbits that
correspond to resonances between the three degrees of freedom.

The structure of phase space in 3-dimensional galactic potentials is
quite complex and we summarize some of its properties here to enable
the reader to more fully appreciate the results of the analysis that
follows. When an integrable potential is perturbed, its phase space
structure is altered, resulting in the appearance of resonances
\citep{lichtenberg_lieberman_92}. Resonances are regions of phase
space where the three fundamental frequencies are not linearly
independent of each other, but two or more of them are related to each
other via integer linear relations. As the perturbation in the
potential increases, the potential deviates further and further from
integrability, and a larger and larger fraction of the phase space
becomes associated with resonances.

In a three dimensional potential, orbits that satisfy one resonance
condition such as $l \omega_x + m\omega_y + n \omega_z = 0$ are
referred to as ``thin orbit'' resonances since they cover the surface
of a two dimensional surface in phase space
\citep{merritt_valluri_99}.  If two independent resonance conditions
between the fundamental frequencies exist, then the orbit is a closed
periodic orbit. Orbits that have frequencies close to the resonant
orbit frequencies are said to be resonantly trapped. Such orbits tend
to have properties similar to that of the parent resonance, but get
``thicker'' as their frequencies move away from the resonance. At the
boundary of the region of phase space occupied by a resonant family is
a region called the ``separatrix''. The separatrix is the boundary
separating orbits with different orbital characteristics.  In this
case it is the region between orbits that have frequencies that are
similar to the resonant orbits and orbits that are not
resonant. Chaotic orbits often occur in a ``stochastic layer'' close
to resonances and at the intersections of resonances. In fact one of
the primary factors leading to an increase in the fraction of chaotic
orbits is the overlap of resonances \citep{chirikov_79}. Chaotic
orbits that are close to a resonant family are referred to as
``resonantly trapped'' or ``sticky orbits''
\citep{habib_kandrup_mahon97} and are often only weakly
chaotic. Orbits that are ``sticky'' behave like the resonant parent
orbit for extremely long times and therefore do not diffuse freely
over their energy surface or undergo significant chaotic mixing.

The frequency analysis method allows one to map the phase space
structure of a distribution function and to easily identify the most
important resonances by plotting ratios of pairs of frequencies
(e.g. $\omega_x/\omega_z$ vs. $\omega_y/\omega_z$) for many thousands
of orbits in the potential. In such a frequency map, resonances appear
as straight lines. Stable resonances appear as filled lines about
which many points cluster, and unstable resonances appear as ``blank"
or depopulated lines. The strength of the resonances can be determined
by the number of orbits that are associated with them.

\subsection{Overcoming microchaos in $N$-body simulations}
\label{sec:chaotic}

 In $N$-body systems like those considered in this paper, the
galactic mass distribution is realised as a discrete set of point
masses.  The discretization of the potential is known to result in exponential
deviation of nearby orbits, even in systems where all orbits are
expected to be regular \citep{miller64, goodman_heggie_hut_93,
kandrup_smith_91, valluri_merritt_00, hemsendorf_merritt_02}.  However
as the number of particles in a simulation is increased, and when
point masses are softened, the majority of orbits begin to appear regular
despite the fact that their non-zero Lyapunov
exponent implies that they are chaotic. \citet{hemsendorf_merritt_02} showed that this Lyapunov
exponent saturates at a finite value beyond a few hundred particles
and corresponds to an $e$-folding timescale of 1/20 of a system
crossing time (for systems with $N \sim 10^5$ particles).  
Despite having large Lyapunov exponents  (i.e. short e-folding
times) these orbits behave and look much like regular orbits
\citep{kandrup_sideris_01, kandrup_siopis_03}. This property of
$N$-body orbits to have non-zero Lyapunov exponents has been referred
to as ``microchaos" \citep{kandrup_sideris_03} or the ``Miller
Instability" \citep{hemsendorf_merritt_02,valluri_etal_07} and
suggests that Lyapunov exponents, while useful in continuous
potentials, are not a good measure of chaotic behavior resulting from
the global potential when applied to $N$-body systems. This is a
strong motivation for our use of a frequency based method which, as we
demonstrate, is extremely effective at distinguishing between regular and
chaotic orbits and is apparently largely unaffected by microchaos.

We now discuss how frequency analysis can be used to distinguish between regular and
chaotic orbits. In realistic galactic potentials most chaotic orbits
are expected to be weakly chaotic and lie close to regular orbits
mimicking their behaviour for long times.  The rate at which weakly
chaotic orbits change their orbital frequencies can be used as a
measure of chaos.  \citet{laskar_93} showed that the change in the
fundamental frequencies over two consecutive time intervals can be
used as a measure of the stochasticity of an orbit. This method has
been used to study the phase space structure in galactic potentials
\citep{pap_las96, pap_las98,valluri_merritt_98}.  Examples of
frequency spectra for each component of motion, and their resolution
into 3 fundamental frequencies are given by \citet{pap_las98} for
different types of orbits. For each time series the spectrum is
analysed and the three fundamental frequencies are obtained. In
Cartesian coordinates the frequencies would be $\omega_x, \omega_y,
\omega_z$.
 
For each orbit we therefore divide the integration time of 50 Gyr time
into two consecutive segments and use NAFF to compute the fundamental
frequencies $\omega_x, \omega_y, \omega_z$ (note that all frequencies
in this paper are in units of Gyr$^{-1}$, therefore units are not
explicitly specified everywhere). We compute the three fundamental
frequencies $\omega_x(t_1), \omega_y(t_1), \omega_z(t_1)$ and
$\omega_x(t_2), \omega_y(t_2), \omega_z(t_2)$ in each of the two
intervals  $t_1$ and $t_2$ respectively. We compute the
``frequency drift'' for each frequency component as:
\begin{eqnarray}
\log(\Delta f_x) = \log{\vert{\frac{\omega_x(t_1)-\omega_x(t_2)}{\omega_x(t_1)}}\vert},\\
\log(\Delta f_y) = \log{\vert{\frac{\omega_y(t_1)-\omega_y(t_2)}{\omega_y(t_1)}}\vert},\\
\log(\Delta f_z) = \log{\vert{\frac{\omega_z(t_1)-\omega_z(t_2)}{\omega_z(t_1)}}\vert}.
\end{eqnarray}
We define the frequency drift parameter $\log(\Delta f)$ (logarithm to base 10) to be the
value associated with the largest of the three frequencies $f_x, f_y,
f_z$ . The larger the value of the frequency drift parameter, the more
chaotic the orbit.

Identifying truly chaotic behavior however also requires that we
properly account for numerical noise.  In previous studies orbits were
integrated with high numerical precision for at least 100 orbital
periods, resulting in highly accurate frequency determination. For
instance, \citet{valluri_merritt_98} found that orbital frequencies in
a triaxial potential could be recovered with an accuracy of $10^{-10}$
for regular orbits and $10^{-4}-10^{-6}$ for stochastic orbits using
integration times of at least 50 orbital periods per
orbit.

In order to use frequency analysis to characterise orbits as regular
or chaotic in $N$-body systems, it is necessary to assess the
numerical accuracy of orbital frequencies obtained by the NAFF
code. To quantify the magnitude of frequency drift that arises purely
from discretization effects (the microchaos discussed above) we select
a system that is spherically symmetric and in dynamical equilibrium.
All orbits in a smooth spherically symmetric potential are rosettes
confined to a single plane \citep{BT08} and are regular. Hence any
drift in orbital frequencies can be attributed entirely to
discretization errors (including minute deviations of the $N$-body
potential from perfect sphericity).  As a test of our application of
the NAFF code to $N$-body potentials we analyse orbits in spherical
NFW halos of two different concentrations ($c=10$ and $c=20$).  The
halos are represented by $10^6$ particles and have mass $\sim 2 \times
10^{12}$ \msun.  Particles in both cases come in two species with
softening of 0.1 kpc and 0.5 kpc.  We carried out the frequency
analysis of 1000 randomly selected orbits which were integrated for 50
Gyr in the frozen $N$-body realisations of each of the NFW halos.
  
\begin{figure}
\includegraphics[width=8.5cm]{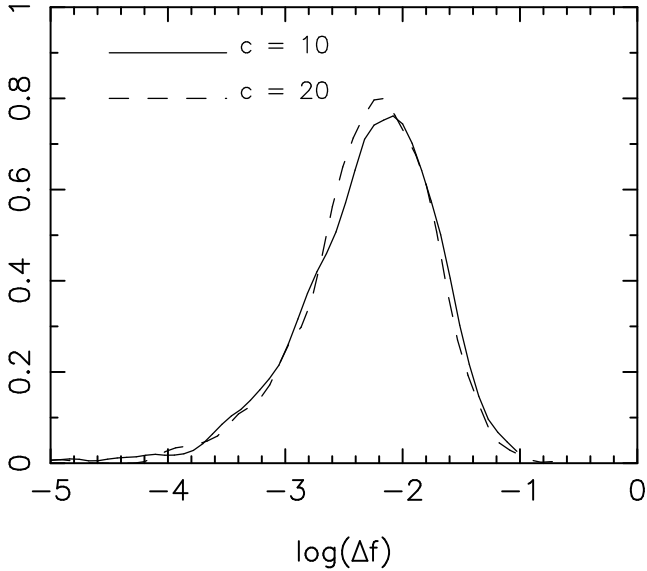}
\caption{Distributions of frequency drift parameter
$\log(\Delta f)$ for 1000 orbits in two different spherical NFW
halos. Despite the difference in concentration $c= 10$ and $c=20$ the
distributions are almost identical having a mean of $\log(\Delta
f) = -2.29$ and a standard deviation $\sigma \simeq 0.56$, with a
significant skewness toward small values of $\log(\Delta f)$.
\label{fig:NFW_hist}}
\end{figure}

Figure~\ref{fig:NFW_hist} shows the distribution of values of
$\log(\Delta f)$ for both halos. In both cases the distribution has a
mean value of $\log(\Delta f) = -2.29$, with standard deviations of
0.58 (for the $c=10$ halo) and 0.54 (for the $c=20$ halo). Both
distributions are significantly skewed toward small values of
$\log(\Delta f)$ (skewness = -0.85) and are more peaky than Gaussian
(kurtosis = 1.95).  Despite the fact that the two NFW halos have
different concentrations, the distributions of $\log(\Delta f)$ are
almost identical, indicating that our chaotic measure is largely
independent of the central concentration.

To define a threshold value of $\log(\Delta f)$ at which orbits are
classified as chaotic we note that 99.5\% of the orbits have values of
$\log(\Delta f) < -1.0$. Since all orbits in a stationary spherical
halo are expected to be regular, we attribute all larger values of
$\log(\Delta f)$ to numerical noise arising from the discretization of
the potential.  Henceforth, we classify an orbit in our $N$-body
simulations to be regular if it has $\log(\Delta f) < -1.0$.

\begin{figure}
\includegraphics[width=8.5cm]{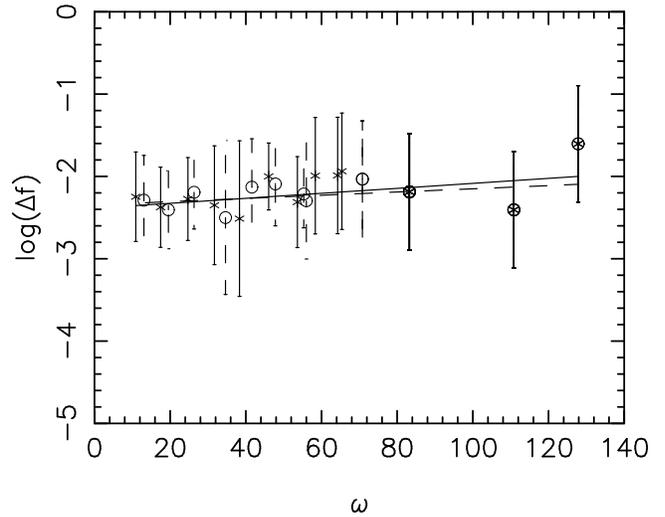}
\caption{$\log(\Delta f)$ versus $\omega$ (in units of Gyr$^{-1}$) for
orbits with $n_p>20$ in the two spherical NFW halos.  Stars are for the halo with
$c=10$ and the open circles are for the halo with $c=20$. The 1000
particles are binned in $\omega$ so that each bin contains the same
number of particles. The vertical error bars represent the standard
deviation in each bin. The straight lines are fits to the data, the
slopes of both lines are consistent with zero.
\label{fig:NFW_df_omega}}
\end{figure}

To accurately measure the frequency of an orbit it is necessary to
sample a significant part of its phase-space structure (i.e. the
surface of a 2-torus in a spherical potential or the surface of the
3-torus in a triaxial potential).  \citet{valluri_merritt_98} showed
that the accuracy of the frequency analysis decreases significantly
when orbits were integrated for less than 20 oscillation
periods. Inaccurate frequency determination could result in
misclassifying orbits as chaotic (since inaccurate frequency
measurement can also lead to larger frequency drifts).  We test the
dependence of $\log(\Delta f)$ on the number of orbital periods $n_p$
by plotting the frequency drift parameter against the largest orbital
frequency (for orbits with $n_p>20$) in both NFW halos in Figure
\ref{fig:NFW_df_omega}\footnote{We use the fractional change in the
largest of the three fundamental frequencies measured over two
contiguous time intervals (frequency drift) as a measure of chaos
\citep{laskar_90}. For situations where a large fraction of orbits is
resonant, it may be more appropriate to use the smallest of the three
frequencies or the component with the largest amplitude.}. We use
$\omega$ instead of $n_p$ since $n_p \propto \omega$ but is harder to
compute accurately.  Particles are binned in equal intervals in
$\omega$ and the error bars represent the standard deviation in each
bin. The straight-lines are best fits to the data-points.  The slopes
of the correlation for the $c=10$ halo (solid line) and for the $c=20$
halo (dot-dashed line) are both consistent with zero, indicating that
$\log(\Delta f)$ is largely independent of $\omega$ (and hence of
$n_p$).  Henceforth we only use orbits which execute more than 20
orbital periods in the 50 Gyr over which they are integrated. The
excluded orbits lie predominantly at large radii and are not
significantly influenced by the changes in the inner halo that are
investigated here. This rejection criterion affects about 25\% of the
orbits in the triaxial dark matter halos that we consider later.

\begin{figure*}
\includegraphics[width=16cm]{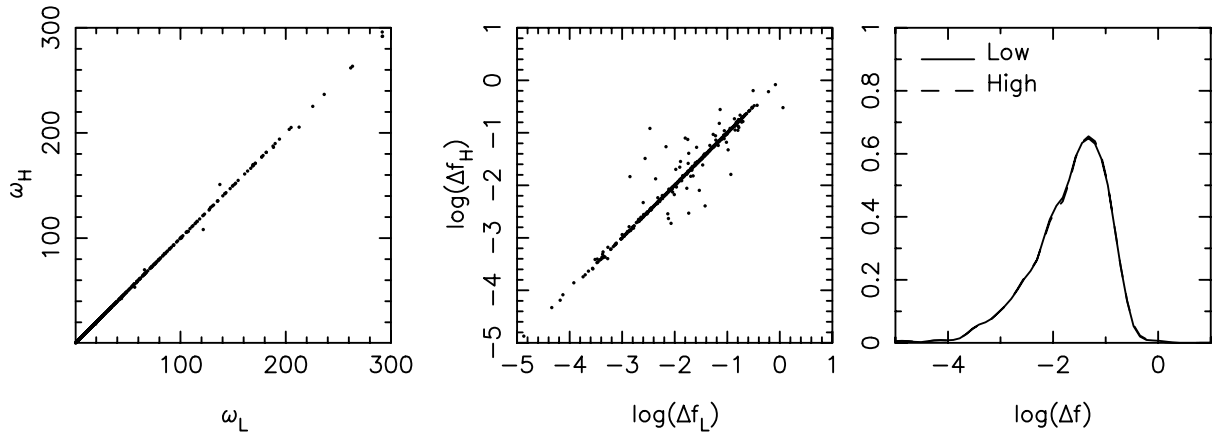}
\caption{{\it left}: comparison of frequencies computed from the low
and high time-sampling runs $\omega_H$ versus $\omega_L$ respectively;
{\it middle}: comparison of diffusion parameter $\log(\Delta f)$
measured in low and high sampling runs; {\it right}: histograms of
diffusion parameter $\log(\Delta f)$.
\label{fig:orSA1bHi_Lo}}
\end{figure*}

The effect of central concentration on the accuracy of frequency
estimation is of particular concern during {\it phase b}, when the
potential is deepened due to the growth of a baryonic component. In
this phase, frequencies of those orbits which are strongly influenced
by the deepened potential are increased. Consequently some orbits
execute many more orbital periods during {\it phase b} than they do in
{\it phase a} or {\it phase c}. However we have fixed the orbital
sampling time period ({\it not integration timestep}) to 1 Myr in all
phases. In principle coarse time sampling should not be a concern
since the long integration time can still ensure a proper coverage of
the phase-space torus. To ensure that the sampling frequency per
orbital period does not significantly alter the frequency estimation
we re-simulated one model (SA1) in {\it phase b} and stored the orbits
5 times more often (i.e. at time intervals of 0.2 Myr). We compared
the frequencies of orbits computed for the low ($\omega_L$) and high
($\omega_H$) time-sampling runs. Figure~\ref{fig:orSA1bHi_Lo} ({\it
left}) shows that there is a strong correspondence between frequencies
obtained with the two different samplings. We also found
(Fig.~\ref{fig:orSA1bHi_Lo} {\it middle}) that the frequency drift
parameter $\log(\Delta f)$ obtained from the two runs are highly
correlated although there is some increase in scatter for orbits with
values of $\log(\Delta f)> -2.$ Since the scattered points lie roughly
uniformly above and below the 1:1 correlation line, there is no
evidence that the higher sampling rate gives more accurate
frequencies.  The right panel shows that the overall distribution of
$\log(\Delta f)$ is identical for the two runs. We find that 95\% of
the particles showed a frequency difference $ < 0.1$\% between the two
different sampling rates. From these tests we conclude that our choice
of sampling rate in the {\it phase b} is unlikely to significantly
affect the frequency measurements of the majority of orbits. 
 We therefore adopt the lower orbit sampling frequency for all
the analysis that follows.

\subsection{Orbit classification}
\label{sec:class}

\citet[][hereafter CA98]{car_agu_98} showed that once a frequency
spectrum of an orbit is decomposed into its fundamental frequencies,
the relationships between the values of the frequencies ($\omega_x,
\omega_y, \omega_z$) can be used to classify the orbits in a triaxial
potential into the major orbit families as boxes, long ($x$) axis
tubes and short ($z$) axis tubes. (CA98 point out that it is difficult
to distinguish between the inner long-axis tubes and outer long-axis
tubes from their frequencies alone.  Therefore we do not attempt to
distinguish between these two families with our automatic
classification scheme.) In addition to classifying orbits into these
three broad categories, they showed that if one or more of the
fundamental frequencies is an integer linear combination of the other
frequencies, the orbit can be shown to be resonant (either a periodic
orbit or an open resonance). We followed the scheme outlined by CA98
to develop our own algorithm to classify orbits as boxes, long-axis
tubes (abbreviated as L-tubes) and short-axis-tubes (abbreviated as
S-tubes) and to also identify orbits that are associated with
low-order resonances.  We do not describe the classification scheme
here since it is essentially identical to that described by CA98, the
main difference lies in that we use NAFF to obtain the fundamental
frequencies of orbits in the $N$-body model, whereas they used a
method based on that of \citet{binney_spergel_84}. We tested our
automated classification by visually classifying 60 orbits that were
randomly selected from the different models. We then ran our automated
orbit classifier on this sample, and compared our visual
classification with that resulting from the automated classifier. The
two methods agreed for 58/60 orbits (a 96\% accuracy rate assuming
that the visual classification is perfectly accurate). Hereafter we assume that our automated
classification is accurate 96\% of the time and therefore any orbit
fractions quoted have an error of $\pm 4$\%.

\subsection{Quantifying orbital shapes}
\label{sec:shapedef}

In any self-consistent potential the distribution of shapes of the
majority of the orbits match the overall shape of the density
profile. The elongation along the major axis is provided either by box
orbits or by inner L-tubes.  The ratios of the fundamental frequencies
of orbits can be used to characterise their overall shape. Consider a
triaxial potential in which the semi-major axis (along the $x$-axis)
has a length $a_x$, the semi-intermediate axis has length $a_y$, and
the semi-minor axis has a length $a_z$. The fact that $a_x > a_y >
a_z$ implies that the oscillation frequencies along these axes are $
|\omega_x| < |\omega_y| < |\omega_z|$ for any (non resonant) orbit
with the same over-all shape as the density distribution (we consider
only the absolute values of the frequencies since their signs only
signify the sense of oscillation).  We can use this property to define
an average ``orbit shape parameter'' ($\chi_s$) for any orbit.  For an
orbit whose overall shape matches the shape of the potential,
 \begin{eqnarray}
  & & |\omega_z| >   |\omega_y| >   |\omega_x| \nonumber \\
  &\Rightarrow&  {\frac{|\omega_y|}{|\omega_z|}}   > {\frac{|\omega_x|}{|\omega_z|}}\nonumber \\
   \chi_s  &\equiv& {\frac{|\omega_y|}{|\omega_z|}} - {\frac{|\omega_x|}{|\omega_z|}} >  0.
 \end{eqnarray}
The orbit shape parameter $\chi_s$ is positive for orbits with
elongation along the figure. The larger the value of $\chi_s$, the
greater the degree of elongation along the major axis.  Very close to
the centre of the potential it is possible for orbits to have greater
extent along the $y$ axis than along the $x$ axis, as is sometimes the
case with outer L-tubes. For such orbits $\chi_s$ is slightly
negative. An orbit for which all frequencies are almost equal would
enclose a volume that is almost spherical. For such an orbit, $\chi_s
\sim 0$ (which we refer to as ``round'').  Note that orbits which are
close to axisymmetric about the short ($z$) axis (i.e. the S-tubes)
also have $\chi_s \sim 0$ because $\omega_x \sim \omega_y$ regardless
of the value of $\omega_z$. Our definition of shape parameter does not
permit us to distinguish between truly round orbits for which
$\omega_x \sim \omega_y \sim \omega_z$ and S-tubes, but both
contribute to a more oblate axisymmetric potential.


\section{Results}
\label{sec:results}

For every model in Table~\ref{tab:simulations} the three fundamental
frequencies $\omega_x, \omega_y, \omega_z$ of each of the orbits in a
selected subsample were computed separately in each of the three
phases. For each orbit the largest of the three fundamental
frequencies is assumed to represent the dominant frequency of
motion. The absolute value of this quantity is referred to as the
largest fundamental frequency: we use $\omega_a, \omega_b, \omega_c$
to refer to the largest fundamental frequencies of an orbit in each of
the three phases {\it a, b, c}. In addition to computing the
fundamental frequencies over the entire 50~Gyr interval, we split the
interval into two equal halves and computed the frequencies in each to
compute the frequency drift parameter $\log(\Delta f)$ defined in \S
\ref{sec:frequency}.  All orbits with $\log(\Delta f) < -1.0$ are
identified as regular and the rest are identified as chaotic. For each
orbit we also compute the total energy ($E$), the absolute value of
the total specific angular momentum ($|j_{\rm tot}|$), the number of
orbital periods ($n_p$), and the pericenter and apocenter distance
from the centre of the potential ($r_{\rm peri}$, $r_{\rm apo}$).

In this section we consider results of five simulations, SA1~\, (\td),
P$_l$A3~\,   (\tb), \PlA4~\,  (\tbbh), P$_f$B2~\,  (\pe), and P$_l$B3~\,  (\pbbh).  We
shall show that halos A and B have very different initial and final
orbital properties despite the fact that their shapes in the presence of baryonic components are
very similar (D08).

\subsection{Distributions of orbital frequencies}
\label{sec:freq_change}

The frequency distribution of randomly selected orbits in a triaxial
halo can be used to characterise the orbital structure of phase space.
It is useful to begin by discussing our expectations for how orbital
frequencies change in response to growth of a central baryonic
component. The potential is significantly deeper in {\it phase b}
compared to {\it phase a}, consequently the most tightly bound orbits
in the initial potential increase their frequencies.  In contrast
orbits which largely lie outside the central mass concentration do not
experience much deformation or much change in their frequencies. The
higher the initial frequency the greater will be the frequency
increase. Hence we expect a faster-than-linear increase in frequency
in {\it phase b} relative to the frequency in {\it phase a}.

\begin{figure*}
\centering
\includegraphics[width=.24\textwidth,angle=-90]{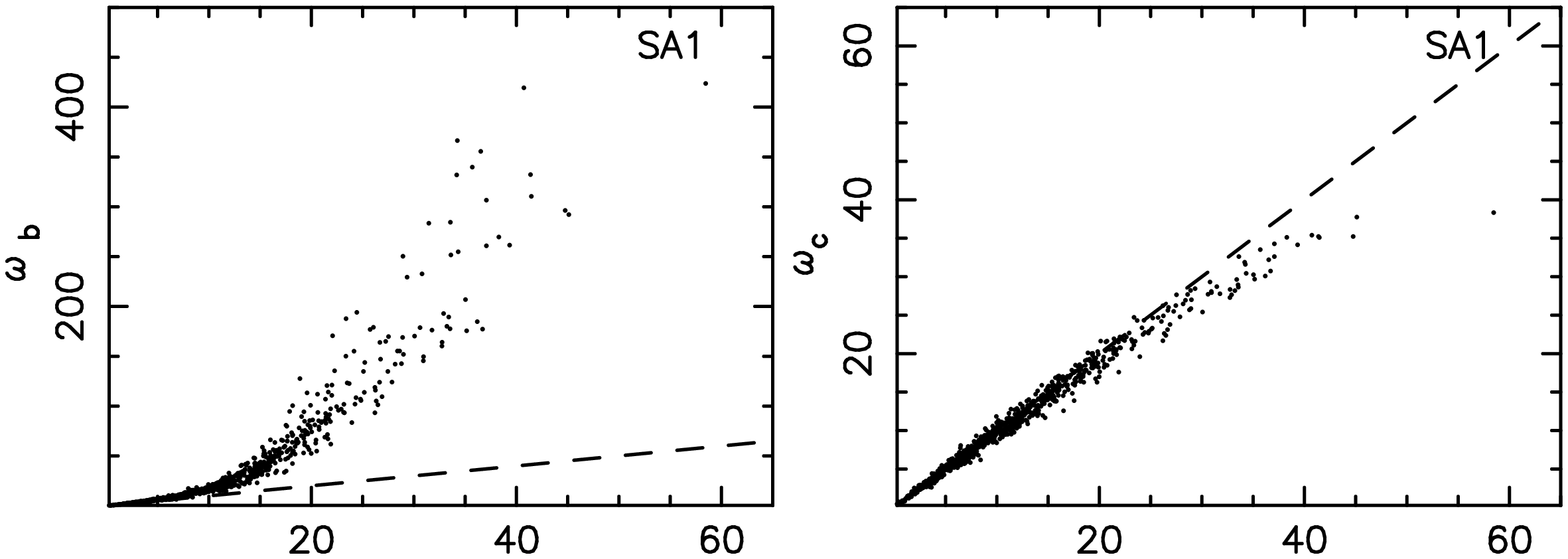}
\includegraphics[width=0.24\textwidth,angle=-90]{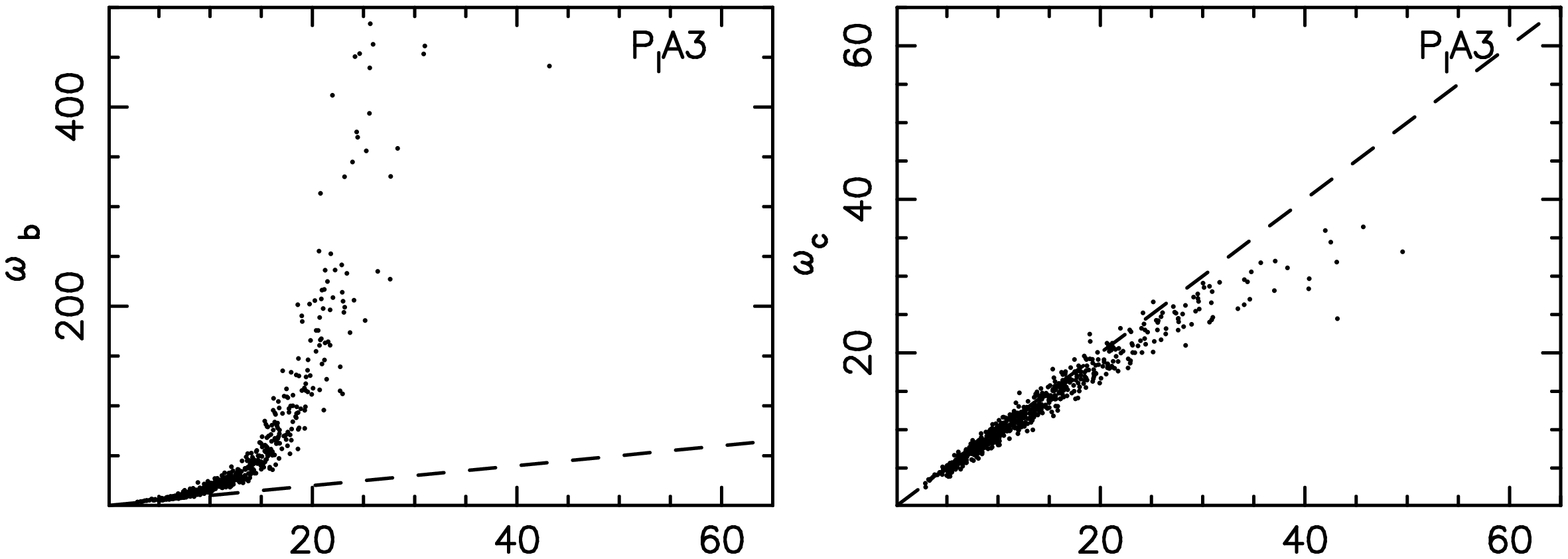}
\includegraphics[width=0.28\textwidth,angle=-90]{orbPfB2_omega_abc.ps}
\caption{For the three models with extended baryonic components the {\it left panels} show
$\omega_b$ versus $\omega_a$ and {\it right panels} show $\omega_c$
versus $\omega_a$ (frequencies in Gyr$^{-1}$). Dashed lines in each panel show the 1:1 correlation
between each pair of frequencies.  From top to bottom the models contain a baryonic disk (SA1), a
spherical bulge (\PlhA3) and a spherical elliptical (\PfB2).
\label{fig:soft_omega_abc}}
\end{figure*}

\begin{figure*}
\centering
\includegraphics[width=0.24\textwidth,angle=-90]{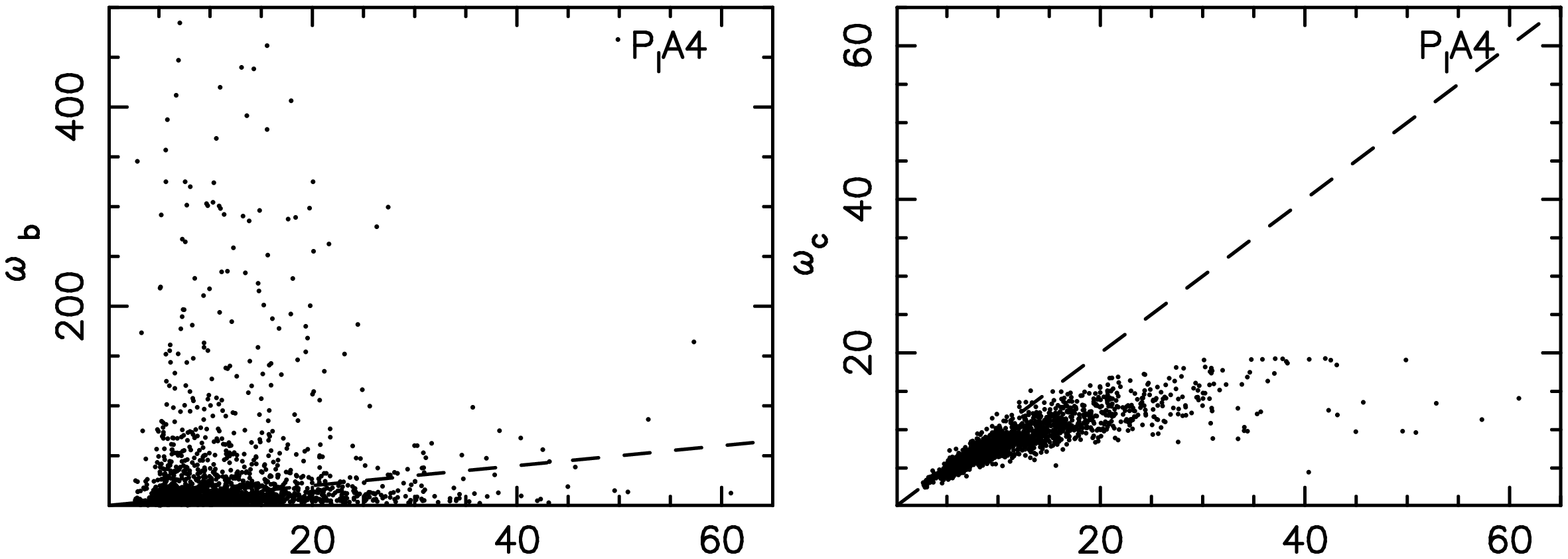}
\includegraphics[width=0.28\textwidth,angle=-90]{orbPlB3_omega_abc.ps}
\caption{Same as Fig.~\ref{fig:soft_omega_abc}, but for the two models with hard point masses of 0.1~kpc softening. Top panels show the effect in the triaxial halo (dominated by box orbits) and bottom panel shows the effect on the prolate halo (dominated by L-tube orbits).
\label{fig:hard_omega_abc}}
\end{figure*}

When the baryonic component is evaporated, the halo expands once more
and the halos regain their triaxiality in models SA1, P$_l$A3 and
P$_f$B2 but are irreversibly deformed in runs P$_l$A4 and P$_l$B3.
One way to investigate the cause of the difference in these behaviours
is to look for correlations between largest fundamental frequencies of
each orbit in each of the three phases.  When the growth of the
baryonic component causes an adiabatic change in orbits, one expects
that their frequencies $\omega_b$ will change in a regular
(i.e. monotonic) way so that the particles which are deepest in the
potential experience the greatest frequency increase.
 In Figures~\ref{fig:soft_omega_abc} and \ref{fig:hard_omega_abc}
we plot correlations between the frequencies $\omega_a$, $\omega_b$
and $\omega_c$.  

In Figures~\ref{fig:soft_omega_abc} we show results for the three
models whose baryonic scale length is greater than 1~kpc: the left
panels show that $\omega_b$ increases faster-than-linearly with
$\omega_a$, as expected with fairly small scatter. The right hand
panels show that $\omega_c$ is quite tightly correlated with
$\omega_a$ in all three models with the tightest correlation for
simulation SA1 (the dashed line shows the 1:1 correlation between the two frequencies). The deviation from the dashed line and the scatter is
only slightly larger in simulations \PlhA3 and \PfB2. The strong
correlation between $\omega_c$ and $\omega_a$ in these models supports
the argument by D08 that the growth of the baryonic component resulted
in regular rather than chaotic evolution. In all three models only a
small fraction of points deviate from the dashed line for the highest
frequencies.

In contrast, the two models with a hard central point mass, \PlB3 and
\PlA4, (Fig.~\ref{fig:hard_omega_abc}) show a non-monotonic change in
$\omega_b$ in response to the growth of the central point mass as well
as a higher degree of scattering in frequency space.  In particular we
note that orbits with small values of $\omega_a$ (i.e. those which are
most weakly bound and have large apocenters) have the largest values
of $\omega_b$, which is in striking contrast to the situation in
Figure~\ref{fig:soft_omega_abc}. We also see that $\omega_b$ sometimes
decreased instead of increasing - again evidence for a scattering in
frequency space rather than an adiabatic change. There is also greater
scatter in the right-hand panels pointing to a less complete recovery
in the frequencies $\omega_c$ after the baryonic component is
evaporated.  Thus we see that when the central point mass is hard and
compact there is significant orbit scattering.

It is clear from a comparison of Figures~~\ref{fig:soft_omega_abc} and
~\ref{fig:hard_omega_abc} that a baryonic component with a scale
length of $R_b \sim 1$~kpc or larger generally causes a regular
adiabatic change in the potential while a hard point mass ($R_b \sim
0.1$~kpc) can produce significant chaotic scattering.

In both Figures~~\ref{fig:soft_omega_abc} and
~\ref{fig:hard_omega_abc} we see in the right hand panels that
$\omega_c < \omega_a$ especially at large values of $\omega_a$
(i.e. all points in the figures lie systematically below the line,
indicating a decrease in the frequency $\omega_c$). Thus particles
must have gained some energy implying that there has been a slight
expansion in the DM distribution following the evaporation of the
baryonic component.

\begin{figure*}
\centering
\includegraphics[width=0.35\textwidth,angle=-90]{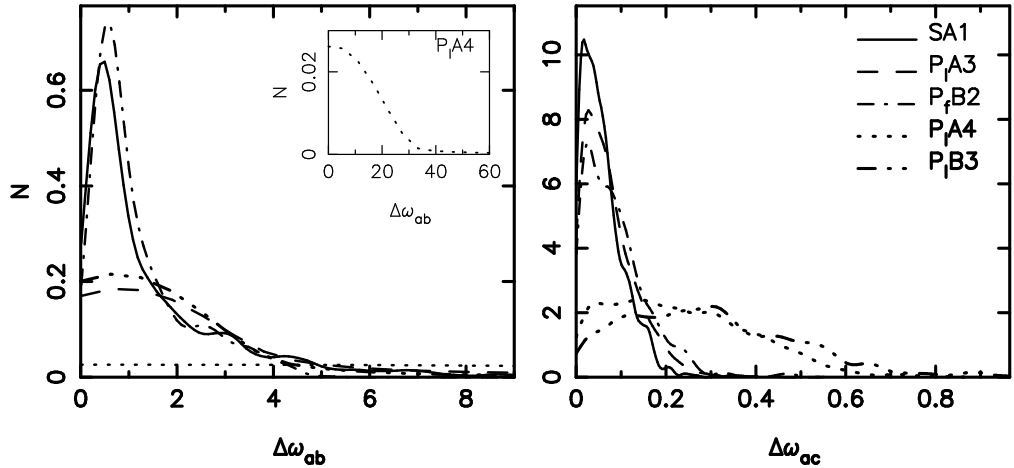}
\caption{Kernel density distributions of $\Delta
\omega_{ab}$ ({\it left panel}) and of $\Delta \omega_{ac}$ ({\it
right panel}) for particles in the halos SA1, P$_f$B2, P$_l$A3, P$_l$A4, and  P$_l$B3
and  as indicated by line legends. Each curve normalised to unit integral. The inset shows the full histogram for P$_l$A4 plotted on a different scale. 
\label{fig:all_omega_change_hist}}
\end{figure*}

What fraction of orbits experience a large fractional change in
frequency of an orbit from {\it phase a} to {\it phase b}, and from
{\it phase a} to {\it phase c}? To investigate this we define:
\begin{eqnarray}
 \Delta \omega_{ab} & = &|(\omega_a - \omega_b)/ \omega_a|\\
\Delta \omega_{ac}  & = & |(\omega_a - \omega_c)/ \omega_a|.
\end{eqnarray}
The first quantity is a measure of the change in frequency
distribution of orbits induced by the presence of the baryonic
component, while the latter quantity measures the irreversibility of
the evolution following the ``evaporation of the baryonic component".
In Figure~\ref{fig:all_omega_change_hist} we plot kernel density
histograms of the distribution of the frequency change $\Delta
\omega_{ab}$ (left panel) and $\Delta \omega_{ac}$ (right panel) for
orbits in all five models as indicated by the line-legends. Each curve is normalized so that the area under it is
unity.

The distribution of $\Delta \omega_{ab}$ is much wider for models
\PlhA3, \PlA4 and \PlB3 than for the other two models. In these three
models the scale-length of the baryonic component is $\leq 1$ kpc and
results in a broad distribution of $\Delta \omega_{ab}$, indicating
that orbits over a wide range of frequencies experience significant
frequency change.  For model P$_l$A4 (dotted line) the histogram of
values of $\Delta \omega_{ab}$ appears almost flat on the scale of this figure because it is spread out over a much larger range of abscissa values indicating that many more particles are significantly scattered in {\it phase b}. The full distribution for P$_l$A4  (see inset panel)  is similar in form to \PlhA3 and \PlB3.

The right panels show that only a small number of orbits in models
SA1, P$_l$A3 and P$_f$B2 experience an irreversible frequency change
$\Delta \omega_{ac} > 20\%$, with the majority of particles
experiencing less than 10\%. In contrast in models P$_l$A4 and
P$_l$B3, the distribution of $\Delta \omega_{ac}$ is much broader: a
significant fraction of particles have experienced a large (20-50\%)
permanent change in their frequencies, reflecting the fact that the
models with a hard-compact point mass are the only ones which
experience irreversible chaotic scattering.

\begin{figure*}
\centering
\includegraphics[width=0.3\textwidth,angle=-90]{orbiSA1_omegach_a_b.ps}
\includegraphics[width=0.3\textwidth,angle=-90]{orbPfB2_omegach_a_b.ps}
\includegraphics[width=0.3\textwidth,angle=-90]{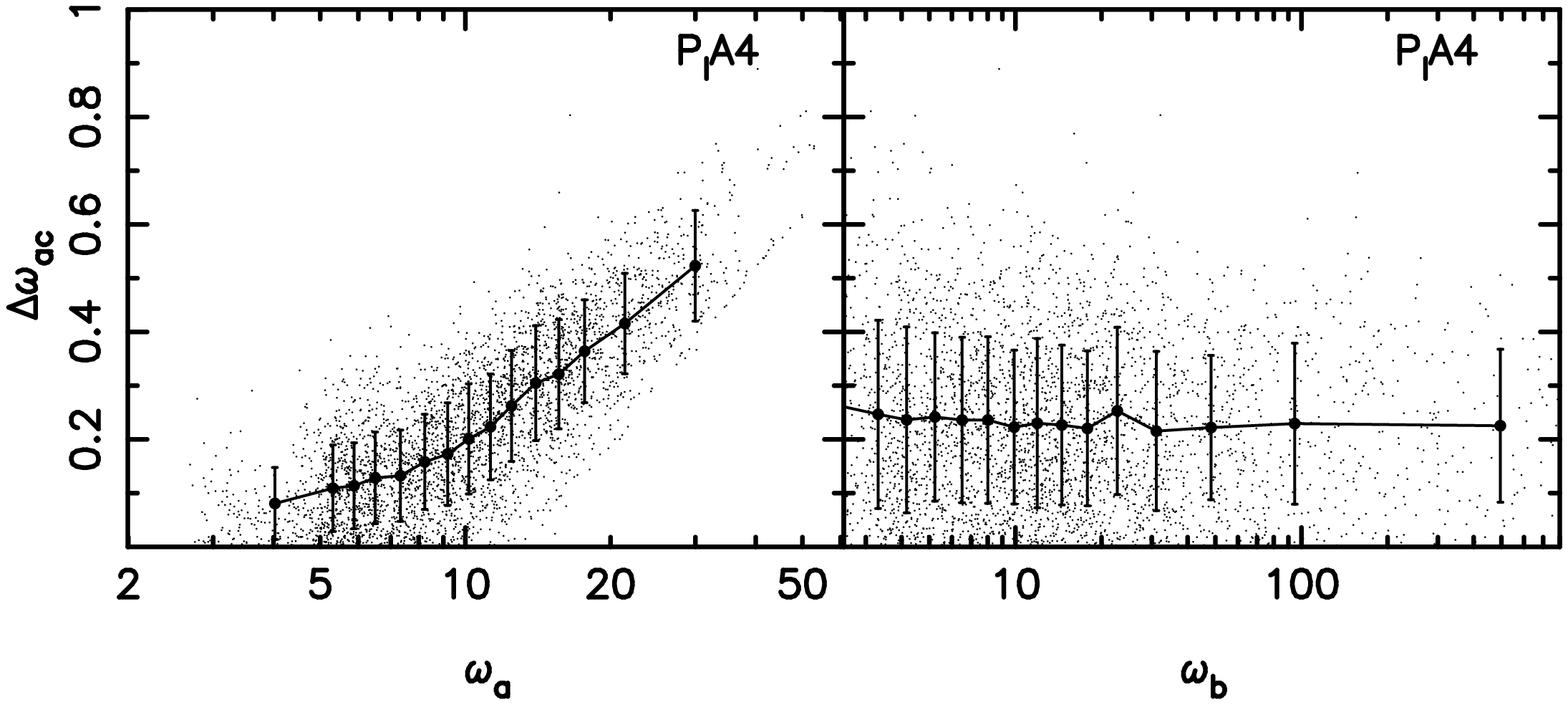}
\includegraphics[width=0.3\textwidth,angle=-90]{orbPlB3_omegach_a_b.ps}
\caption{For four models small dots show $\Delta
\omega_{ac}$ versus $\omega_a$ ({\it left}) and versus $\omega_b$ ({\it
right}) for all particles analysed. Large solid dots with error bars
show mean and standard deviation of particles in 15 bins in frequency.
Top two panels are for models with an extended baryonic component 
(SA1, \PfB2), lower two panels show models (\PlA4, \PlB3) with a compact baryonic component. \label{fig:all_domega_omegaa}}
\end{figure*}

\begin{figure*}
\centering
\includegraphics[width=0.4\textwidth]{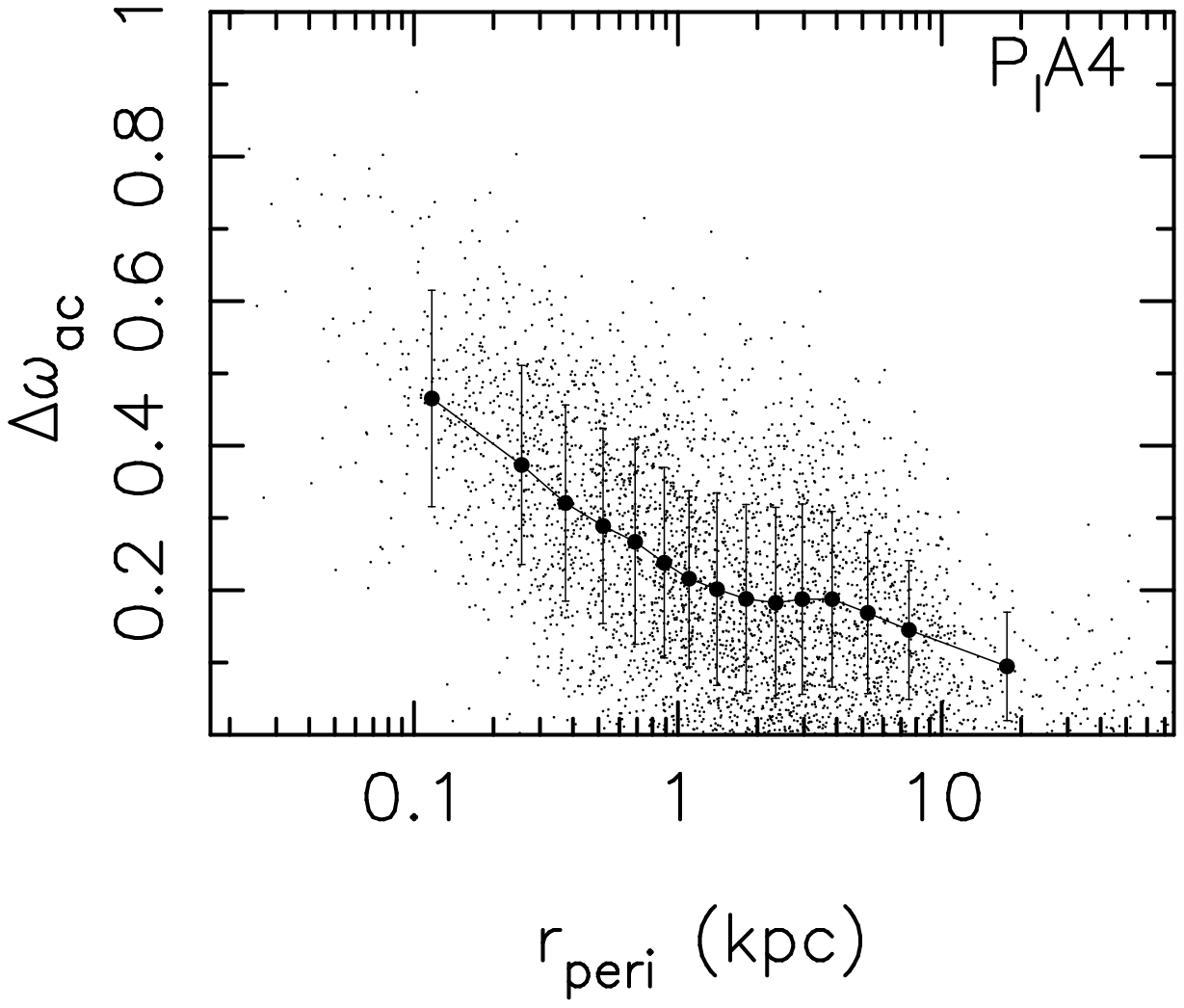}
\includegraphics[width=0.4\textwidth]{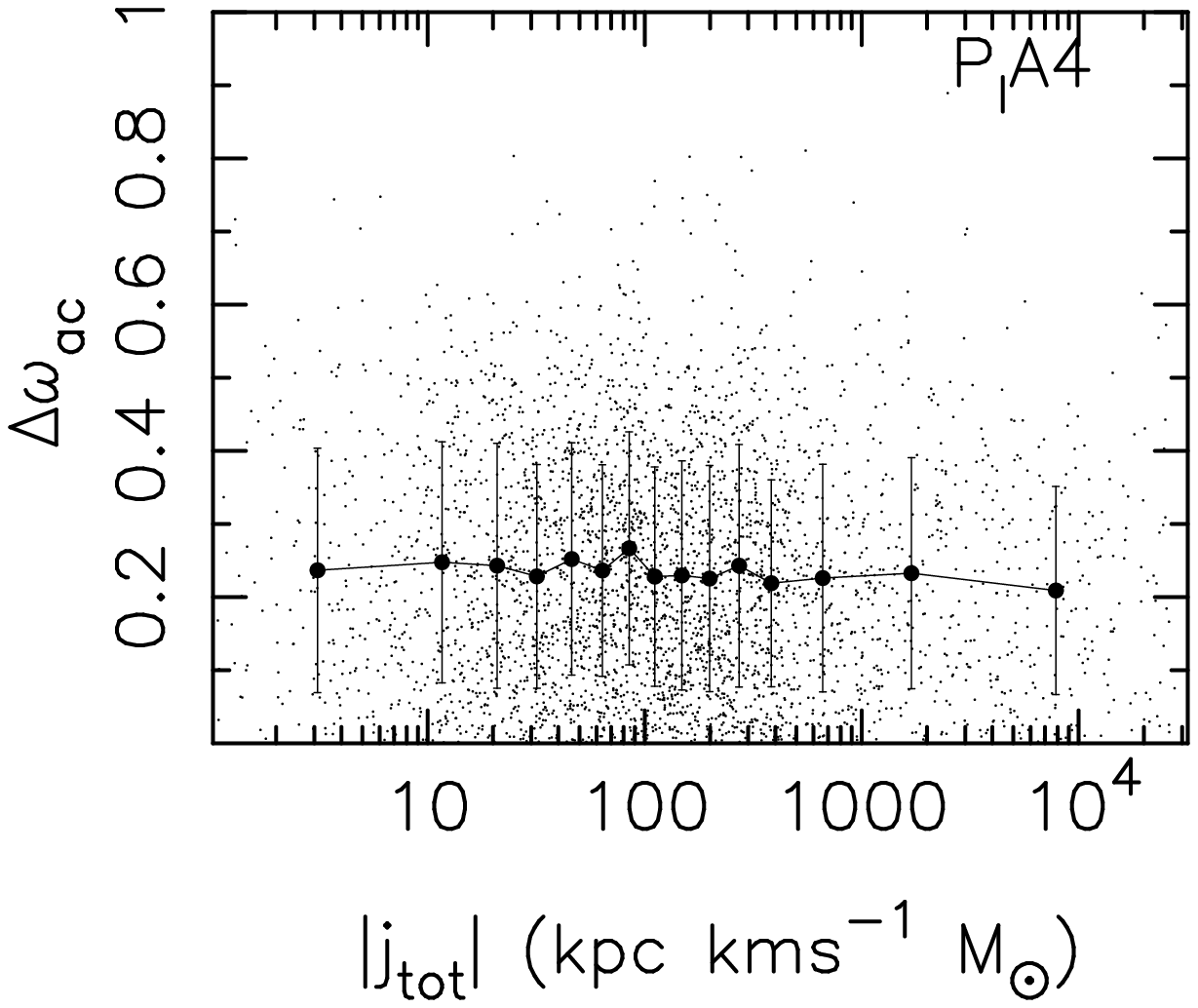}
\includegraphics[width=0.4\textwidth]{orbPlB3_rperi.ps}
\includegraphics[width=0.4\textwidth]{orbPlB3_Jtot.ps}
\caption{For model \PlA4 (top panels) and \PlB3 (bottom panels)
{\it left:} $\Delta
\omega_{ac}$ versus $r_{\rm peri}$; {\it right:} $\Delta \omega_{ac}$
versus $|j_{\rm tot}|$.  
\label{fig:Pl3_jtot_rperi}}
\end{figure*}

Are there specific orbital characteristics that contribute to a large
permanent frequency change, $\Delta \omega_{ac}$, between the two
triaxial phases? We address this by determining how this quantity
relates to other orbital properties.  In
Figure~\ref{fig:all_domega_omegaa} we plot $\Delta \omega_{ac}$ versus
$\omega_a$ ({\it left panels}) and versus $\omega_b$ ({\it right
panels}) for four of our models. In the top two panels (SA1, and \PfB2
- models with an extended baryonic component) there is no evidence of
a dependence of frequency change on $\omega_b$ and only a slight
increase in $\Delta \omega_{ac}$ at the highest values of $\omega_a$
(results for \PlhA3 are not shown but are very similar to those for
SA1).

On the other hand, the lower two panels (\PlA4 and \PlB3 - models with
a compact hard baryonic component) show that there is a strong
correlation between $\Delta \omega_{ac}$ and orbital frequency
$\omega_a$ indicating that the orbits with the highest frequencies
($\omega_a$) experience the largest frequency change, $\Delta
\omega_{ac}$. This is evidence that scattering by the hard central
point mass is greatest for particles that are most tightly bound and
therefore closest to the central potential, confirming previous
expectations \citep{ger_bin_85, merritt_valluri_96}. The absence of an
appreciable correlation with $\omega_b$ is the consequence of
scattering of orbits in frequency.

In Figure~\ref{fig:Pl3_jtot_rperi} we plot $\Delta \omega_{ac}$ versus
$r_{\rm peri}$ ({\it left panels}) and versus $|j_{\rm tot}|$ (the
total specific angular momentum of an orbit averaged over its entire
orbit in {\it phase a}) ({\it right panels}) for orbits in the two
models with compact central point mass (\PlA4 and \PlB3).  (We do not
show plots for models SA1, \PlhA3 and \PfB2, because they show no
correlation between $\Delta \omega_{ac}$ and either $|j_{\rm tot}|$ or
$r_{\rm peri}$.)  The left panels of Figure~\ref{fig:Pl3_jtot_rperi}
shows that orbits which pass closest to the central point mass
experience the most significant scattering. The absence of a
correlation with $|j_{\rm tot}|$ however indicates that scattering
{\em is independent of the angular momentum} of the orbit.  In the
next section we show that halo A (the initial triaxial halo for model
\PlA4) is dominated by box orbits while halo B is initially prolate,
and is dominated by L-tubes which circulate about the long
axis.  Contrary to the prevailing view that centrophilic box orbits
are more strongly scattered by a central point mass than centrophobic
tube orbits, these figures provide striking evidence that chaotic
scattering is equally strong for the centrophobic L-tubes that
dominate model \PlB3 as it is for the centrophilic box orbits that
dominate model \PlA4.
 We return to a fuller discussion of
the cause of this scattering in \S~\ref{sec:summary}.

\subsection{Changes in Orbital Classification}
\label{change_type}

\begin{table*}
\caption{\label{tab:orbit_class} Orbit composition of the models.  The
numbers represent the fraction of orbits in each family.}
\begin{center}
\begin{tabular}{lccccccccccccccc}
\hline
{\bf Type}  &\multicolumn{3}{c}{\bf Run SA1} &\multicolumn{3}{c}{\bf Run \PlhA3}& \multicolumn{3}{c}{\bf  Run \PfB2} &\multicolumn{3}{c}{\bf  Run \PlA4} &\multicolumn{3}{c}{\bf  Run \PlB3}  \\
\hline
\hline
{\it Phase}       &{\it  a}  & {\it  b} & {\it  c}   & {\it  a} & {\it  b} & {\it  c}   & {\it  a} & {\it  b } & {\it  c}   & {\it  a} & {\it  b } & {\it  c}  & {\it  a} & {\it  b } & {\it  c} \\
\hline
{\bf  Boxes}     & 0.86  & 0.43 & 0.83 & 0.84 &  0.16 & 0.76 &  0.15 & 0.09 & 0.29 &  0.84 & 0.17 &0.80  & 0.15 & 0.03  & 0.21    \\
{\bf  L-tubes}   & 0.11  & 0.09 & 0.12 & 0.12 &  0.43 & 0.15 &  0.78 & 0.75 & 0.54 &  0.12 & 0.35 & 0.11 &  0.78 & 0.78 & 0.59    \\
{\bf  S-Tubes} & 0.02  & 0.27 & 0.03 & 0.02 &  0.33 & 0.06 &  0.07 & 0.09 & 0.16 &  0.02 & 0.26 & 0.04 & 0.07  & 0.11 & 0.14    \\
{\bf  Chaotic}   & 0.01  & 0.21 & 0.02 & 0.02 &  0.08 & 0.03 &  0.00 & 0.07 & 0.01 &  0.02 & 0.21 & 0.05 & 0.00 & 0.08  & 0.06    \\
\hline
\end{tabular}
\end{center}
\end{table*}

As we discussed in \S~\ref{sec:class}, relationships between the
fundamental frequencies of a regular orbit can be used to classify it
as a box orbit, a L-tube or a S-tube orbit.  Quantifying the orbital
composition of the two different halos A and B and how their
compositions change in response to the growth of a baryonic component
yields further insight into the factors that lead to halo shape
change.  Orbits were first classified as regular or chaotic based on
their drift parameter $\log(\Delta f)$ as described in
\S~\ref{sec:chaotic}. Regular orbits were then classified into each of
three orbital families using the classification scheme outlined in
\S~\ref{sec:class}. The results of this orbit classification for each
model, in each of the three phases, are given in
Table~\ref{tab:orbit_class}.

The most striking difference between the initial triaxial models is
that halo A ({\it phase a} of models SA1, \PlhA3 and \PlA4) is
dominated by box orbits (84-86\%) while halo B ({\it phase a} of
models \PfB2 and \PlB3) is dominated by L-tubes (78\%).  (The small
differences between models SA1, \PlhA3 and \PlA4 in {\it phase a} is
purely a consequence of the selection of different subsets of orbits
from halo A.)  None of the initial models has a significant fraction
of S-tubes or of chaotic orbits.

The very different orbit compositions of halos A and B in {\it phase
a} results in rather different evolutions of their orbital populations
in response to the growth of a central baryonic component.  Although
the growth of the disk results in a significant decrease in the box
orbit fraction (from 86\% to 43\%) with boxes being converted to
either S-tubes or becoming chaotic in {\it phase b}, model SA1 is
highly reversible suggesting an adiabatic change in the potential. In
model \PlhA3 and \PlA4, the more compact spherical baryonic components
decrease the fraction of box orbits even more dramatically (from 84\%
down to 16-17\%), pointing to the vulnerability of box orbits to
perturbation by a central component. Despite the similar changes in
the orbital populations of the two models, \PlhA3 is much more
reversible than model \PlA4, indicating that both the shape of the
central potential and its compactness play a role in converting box
orbits to other families and that the change in orbit type is not
evidence for chaotic scattering. It is striking that the more compact
point mass in model \PlA4 results in significantly more chaotic orbits
(21\%) compared to 8\% in \PlhA3.

\begin{figure*}
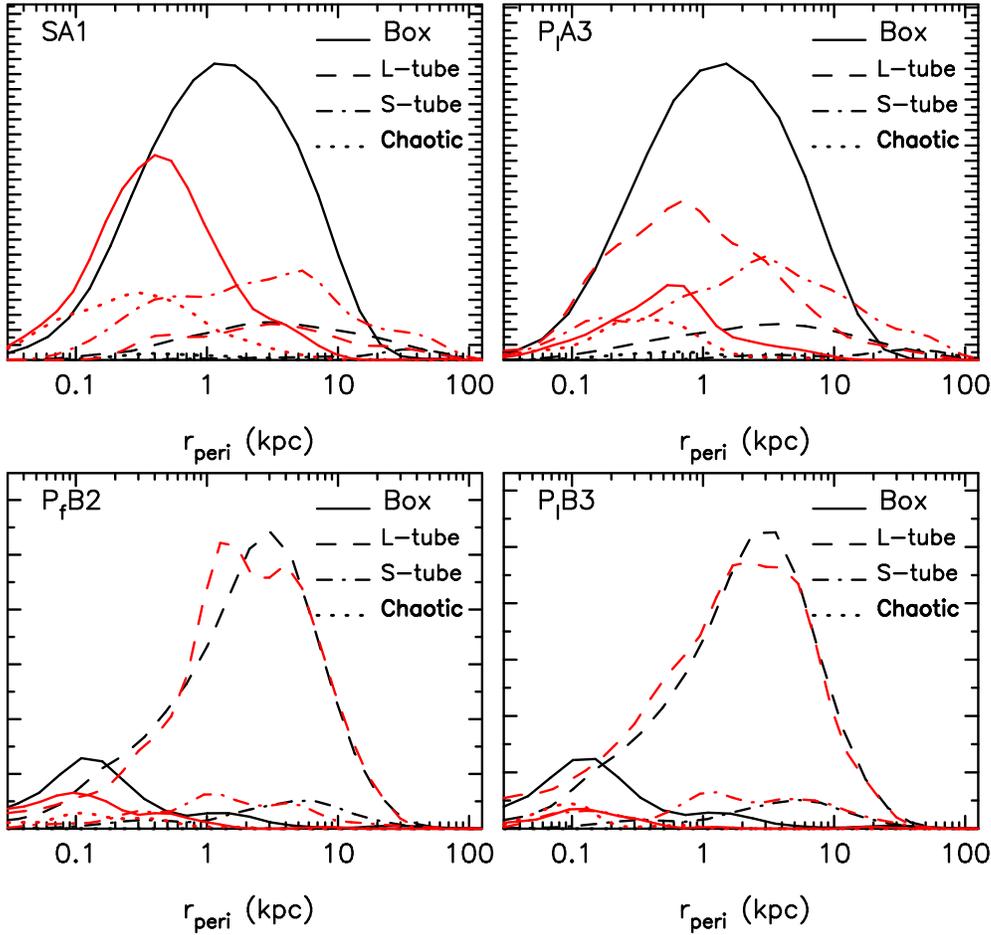

\centering
\includegraphics[width=0.35\textwidth,angle=-90]{orb5SA1_typab_rp.ps}
\includegraphics[width=0.35\textwidth,angle=-90]{orbPlA3_typab_rp.ps}
\includegraphics[width=0.35\textwidth,angle=-90]{orbPfB2_typab_rp.ps}
\includegraphics[width=0.35\textwidth,angle=-90]{orbPlB3_typab_rp.ps}
\caption{Distributions of $r_{\rm peri}$ for different orbit
types. Distributions of each of the four different orbital types as
indicated by the line-legends. Distribution in {\it phase a} is given
by black curves and distribution in {\it phase b} is shown by red
curves. The integral under each curve is proportional to the number of orbits
of that orbital type.  
\label{fig:typab_rp}}
\end{figure*}

While halo A is initial dominated by box orbits, halo B is initially
dominated by L-tubes, which dominate the orbit population in halo B in
all three phases.  The growth of the baryonic component in {\it phase
b} causes the box orbit fraction to decrease (especially in model
\PlB3) while the fraction of chaotic orbits increases slightly. The
more extended point mass in \PfB2 causes a larger fraction of L-tubes
to transform to orbits of another type than does the harder point mass
in \PlB3, despite the fact that there is much greater scattering in
the latter model. A comparison between the model \PlA4 and \PlB3 show
that their orbit populations in the presence of a baryonic component
differ significantly due to the different original orbit populations,
while their degree of irreversibility is identical
(e.g. Fig.~\ref{fig:all_omega_change_hist}) since the point mass in
the two models is identical.

A significant fraction (21\%) of the orbits in {\it phase b} of model
SA1 and \PlA4 are classified as chaotic (orbits with drift rate
$\log(\Delta f) > -1.0$), in comparison with 9\%, 7\% and 8\% in
models \PlhA3, \PfB2 and \PlB3 respectively. While the presence of
such a large fraction of chaotic orbits in {\it phase b} of model
\PlA4 may be anticipated from previous work, the high fraction of
chaotic orbits in SA1 (\td) is puzzling. To address concerns about
classification error that could arise from errors in the accuracy of
our frequency computation, we showed, in Figure~\ref{fig:orSA1bHi_Lo},
that changing the frequency at which orbits were sampled by a factor
of five did not result in any change in the overall distribution of
$\log(\Delta f)$, and hence should not affect our classification of
orbits as regular or chaotic.  Another puzzling fact is that, although
model SA1 in {\it phase b} has such a significant fraction of chaotic
orbits, the orbit fractions essentially revert almost exactly to their
original ratios once the disk is evaporated in {\it phase c}. Hence,
the large fraction of chaotic orbits in {\it phase b} do not appear to
cause much chaotic mixing. We will return to a more complete
investigation of this issue in \S~\ref{sec:maps}.

In Figure~\ref{fig:typab_rp} we investigate how orbits of different
types (boxes, L-tubes, S-tubes, chaotic) are distributed with $r_{\rm
peri}$, and how this distribution changes from {\it phase a} (black
curves) to {\it phase b} (red curves).  In {\it phase a} the initially
triaxial halo A models (black curves) are dominated by box orbits. The
fraction of box orbits is significantly decreased in {\it phase b}. In
particular box orbits with large $r_{\rm peri}$ are transformed
equally to short axis tubes and chaotic orbits, while some box orbits
at small $r_{\rm peri}$ are converted to long-axis tubes. In contrast
halo B models are dominated by L-tubes in both phases.  Rather
striking is how little the fraction of L-tubes in the halo B models
changes, despite the fact that the halos are significantly more oblate
axisymmetric in {\it phase b} than in {\it phase a}.  We saw in
Figure~\ref{fig:all_omega_change_hist} that a significant fraction of
orbits experience strong scattering that manifests as a change in
their orbital frequencies, and in Figure~\ref{fig:Pl3_jtot_rperi} we
noted that the orbits with the smallest pericenter radii experience
the largest change in frequency. In both models \PlA4 and \PlB3 the
compact central point mass significantly reduced the box
orbits. However model \PlB3 only has a small fraction (15\%) of box
orbits and it seems unlikely that the chaotic scattering of this small
fraction of orbits off the central point mass is entirely responsible
for driving the evolution of halo shape. Also \PfB2 which has a much
more extended baryonic component shows a change in orbit population
which closely parallels \PlB3 and we saw that \PfB2 is quite
reversible and shows little evidence for chaotic scattering.  It is
clear (from Fig.~\ref{fig:typab_rp}) that in the prolate models (halo
B) the majority of the orbits are L-tubes with large pericenter radii
($\left<r_{\rm peri}\right> \sim 3$~kpc) and these remain L-tubes in
{\it phase b}. How then do these prolate models evolve to more
spherical models while retaining their dominant orbit populations? To
address this question we will now investigate the distribution of
orbital shapes in each model at each phase of the evolution.

\subsection{Changes in orbital shape}
\label{sec:shape}

\begin{figure*}
\centering
\includegraphics[width=0.4\textwidth]{orbiSA1_shape.ps}
\includegraphics[width=0.4\textwidth]{orbPfB2_shape.ps}
\includegraphics[width=0.4\textwidth]{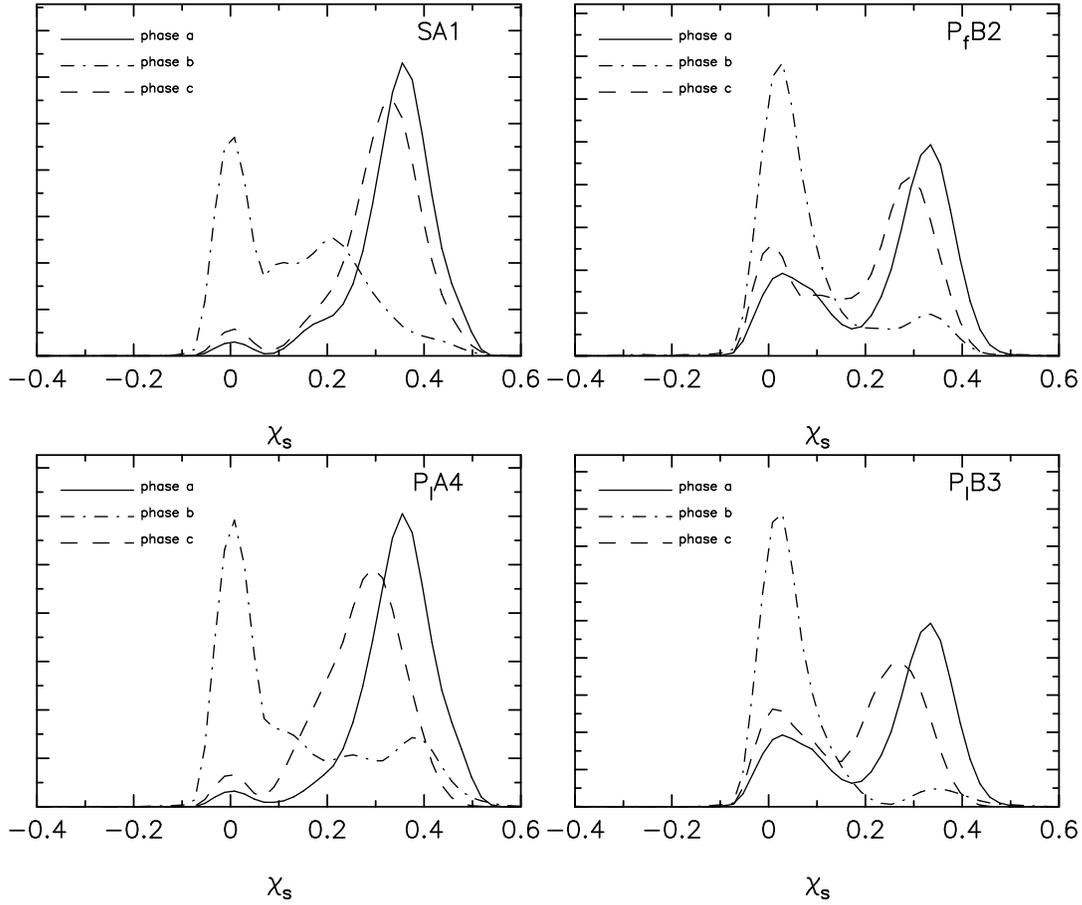}
\includegraphics[width=0.4\textwidth]{orbPlB3_shape.ps}
\caption{Kernel density histograms of the distribution of orbital
shape parameter $\chi_s$ for each of the four models: SA1 ({\it top
left}), \PlA4 {\it bottom left }, \PfB2 ({\it top right }) and \PlB3
({\it bottom right}) (\PlhA3 is not shown since it is very similar to \PlA4.) 
Distributions of $\chi_s$ in {\it phase a} are
shown by solid curves, in {\it phase b} by dot-dashed curves, and in
{\it phase c} by dashed curves. In all models, a large fraction of
orbits in {\it phase b} are ``round'' ($\chi_s \simeq 0.$.)
\label{fig:all_shapes}}
\end{figure*}

A parameter, $\chi_s$, to quantify the shape of an orbit was defined
in Equation~3 of \S~\ref{sec:shapedef}. Recall that this quantity is
positive when the orbit is elongated along the major axis of the
triaxial figure, is negative when elongated along the intermediate
axis, and almost zero when the orbit is ``round'' ($\omega_x \sim
\omega_y \sim \omega_z$) or roughly axisymmetric about the minor axis
($\omega_x \sim \omega_y$). In Figure~\ref{fig:all_shapes} we show the
shape distributions for the orbits in four of our five models. For
each model we show kernel density histograms for models in {\it phase
  a} (solid curves), {\it phase b} (dot-dashed curve), and {\it phase
  c} (dashed curves). In each plot the curves are normalized such that
the integral under each curve is unity. We define orbits to be
elongated if $\chi_s \ga 0.25$, and to be ``round'' if $|\chi_s| \le
0.1$.
 
Before the growth of the baryonic component ({\it phase a}: solid
curves) the halo A models (left panels SA1, \PlA4) have a distribution
of orbital shapes that has a large peak at $\chi_s \sim 0.35$, arising
from elongated orbits and a very small peak at $\chi_s \sim 0$ due to
round orbits (model \PlhA3 is not shown but is similar to
\PlA4.)  In halo B models, (right panels \PfB2, \PlB3) on the other
hand, the distribution of shapes is double peaked with about one third
of all orbits contributing to the peak at $\chi_s \sim 0$. This
implies one third of its orbits in the initially prolate halo B are
``round''.  In both halo A and B however, the larger of the two peaks
has a value of $\chi_s \sim 0.35$ corresponding to quite elongated
orbits.  Despite the quite different underlying orbital distributions
(halo A models dominated by box orbits while the halo B models are
dominated by L-tubes).  This illustrates that despite having different
orbital compositions, a significant fraction of their orbits are
similarly elongated.

The dot-dashed curves in all the panels show the distribution of
orbital shapes in {\it phase b}. In all four models there is a
dramatic increase in the peak at $|\chi_s| \sim 0$, pointing to a
large increase in the fraction of round (or S-tubes) at the expense of
the elongated (L-tube or box) orbits.  In the halo B models the
elongated orbits are significantly diminished indicating that the
elongated L-tubes in {\it phase a} are easily deformed to ``round"
orbits in {\it phase b} (most likely squat inner-L-tubes).  However,
in model SA1 there is a large fraction of orbits with intermediate
values of elongation $0.1 \leq \chi_s \leq 0.4$.

In {\it phase c} (dashed curves) all models show the dominant peak
shifting back to quite high elongation values of $\chi_s \sim 0.3$
(although this is slightly lower than $\chi_s \sim
0.35$ in {\it phase a}). The downward shift in the peak is most evident in model \PlB3
(\pbbh), which as we saw before, exhibits the greatest irreversibility
in shape. The scattering of a large fraction of the orbits by the hard
central potential in model \PlB3 seen in
Figures~\ref{fig:all_domega_omegaa} and \ref{fig:Pl3_jtot_rperi} is
the major factor limiting reversibility of the potential. The smallest
shift is for model SA1 (\td), which exhibited the greatest
reversibility.

\begin{figure*}
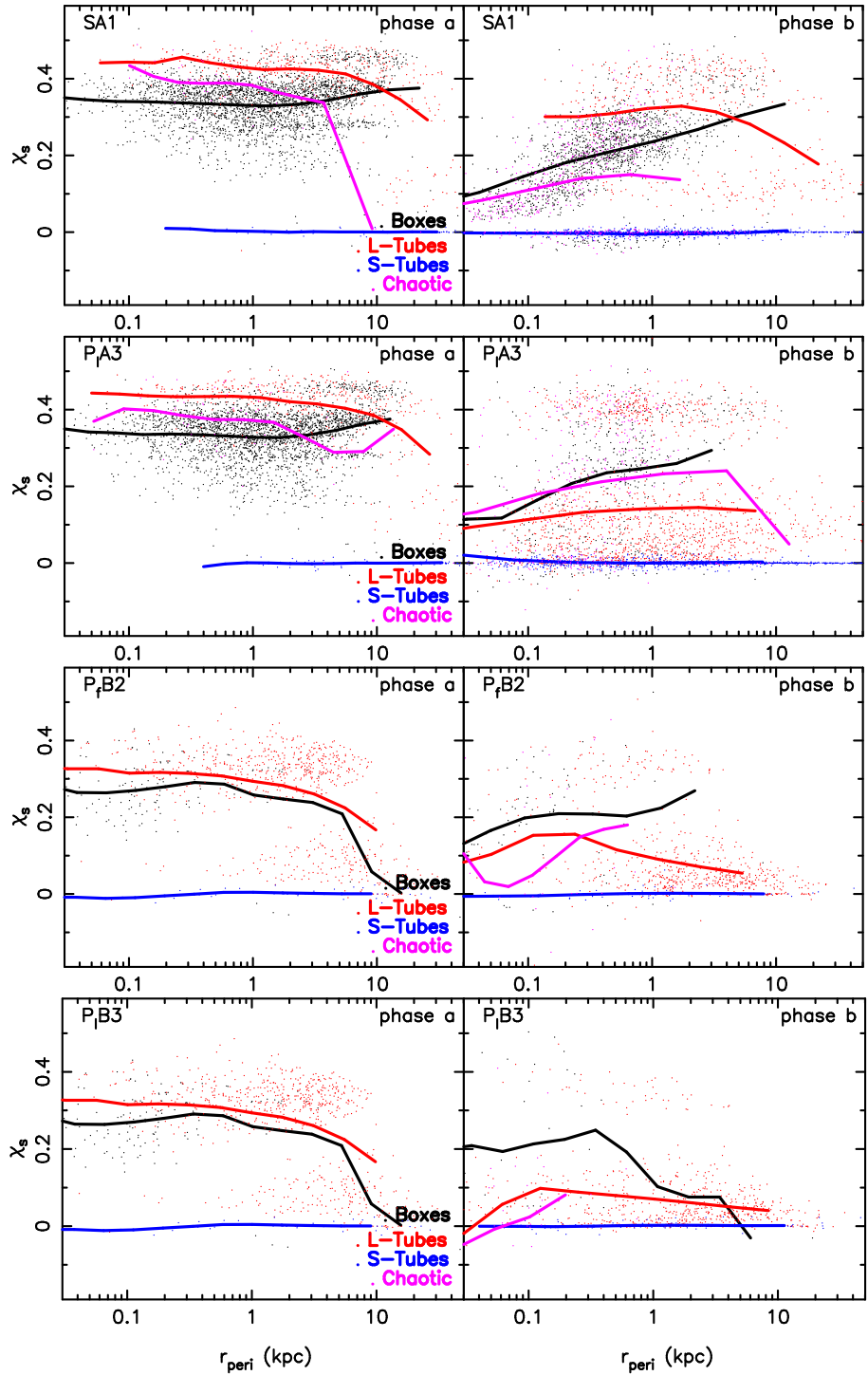

\centering
\includegraphics[width=0.26\textwidth,angle=-90]{orb5SA1_sh_rp.ps}
\includegraphics[width=0.26\textwidth,angle=-90]{orbPlA3_sh_rp.ps}
\includegraphics[width=0.26\textwidth,angle=-90]{orbPfB2_sh_rp.ps}
\includegraphics[width=0.30\textwidth,angle=-90]{orbPlB3_sh_rp.ps}
\caption{For models SA1, \PlhA3, \PfB2 and \PlB3, the
orbital shape parameter $\chi_s$ for each orbit is plotted against its
pericentric radius $r_{\rm peri}$ as a small dot. The orbits of each
of the four major orbital families are colour coded as in the figure
legends. Left hand panels are for {\it phase a} and
right hand panels are for {\it phase b}.  The solid curves show the
mean value of $\chi_s$ for all particles of that particular family, in
15 bins in $r_{\rm peri}$. Curves are not plotted if there are fewer
than 30 orbits in a given orbital family.  We used a kernel regression
algorithm to smooth the curves.
\label{fig:shape_peri}}
\end{figure*}

We can also investigate how the shapes of orbits vary with pericentric
radius.  We expect that orbits closer to the central potential should
become rounder ($\chi_s \rightarrow 0$) than orbits further out. We
see that this expectation is borne out in Figure~\ref{fig:shape_peri}
where we plot orbital shape parameter $\chi_s$ versus $r_{\rm peri}$
in both {\it phase a} (left hand plots) and {\it phase b} (right hand
plots).  In each plot the dots show values for individual orbits. The
solid curves show the mean of the distribution of points in each of 15
bins in $r_{\rm peri}$. Curves are only plotted if there are more than
30 particles in a particular orbital family (\PlA4 is not shown since
it is similar to \PlhA3).  For models SA1 (\td) and \PlhA3 (\tb) the
figure confirms that elongated orbits in the initial halo A were box
orbits (black dots and curves) and L-tubes (red dots and
curves). The S-tubes (blue dots and curves) are primarily
responsible for the ``round'' population at $\chi_s \sim 0$.  In {\it
phase b} (right-hand panels) of both SA1 and \PlhA3 there is a clear
tendency for the elongated orbits (boxes, L-tubes and chaotic)
to become rounder at small pericenter distances, but they continue to
be somewhat elongated at intermediate to large radii.  Chaotic orbits
in {\it phase b} of model SA1 appear to span the full range of
pericentric radii and are not confined to small radii. (Note that the
density of dots of a given colour is indicative of the number of
orbits of a given type but the relative fractions are better judged
from Fig.~\ref{fig:typab_rp} and Table~2.)

For {\it phase a} in the models \PfB2 (\pe) and \PlB3 (\pbbh) (left
panels of each plot), boxes and L-tubes are elongated ($\chi_s \ge
0.25$), except at large $r_{\rm peri} \ge 8$~kpc where they become
rounder.  We see a trend for the average orbital
shape (as indicated by the curves) in {\it phase b} to become round at small pericenter radii. 

Note that in all the plots the curves only show the average shape of orbits of a given type at any radius. The points show that in the case of the L-tubes in particular, the red dots tend to be distributed in two ``clouds": one with large elongations $\chi_s > 0.3$ and one with small elongation $\chi_s \sim 0.1$. 

Thus in all four models it is clear that orbits that are elongated
along the major axis of the triaxial potential in {\it phase a} become
preferentially rounder at small pericenter radii in {\it phase b}. It
is this change in orbital shape that plays the most significant role
in causing the overall change in the shape of the density in the
baryonic phase\footnote{Due to our chosen definition of shape
parameter, S-tubes generally have $\chi_s \sim 0$ regardless of radius,
because $\omega_x \sim \omega_y$.}.

\subsection{Frequency maps and chaotic orbits}
 \label{sec:maps}

We saw in Table~\ref{tab:orbit_class} that {\it phase b} of model SA1
(\td) and of model \PlA4 (\tbbh) have a significant fraction (21\%) of
chaotic orbits (i.e. orbits with $\log(\Delta f) > -1$).  While \PlA4
shows significant lack of reversibility, which we can attribute to the
presence of this high fraction of chaotic orbits, model SA1 does not
show evidence for irreversibility.

\begin{figure}
\centering
\includegraphics[width=0.4\textwidth]{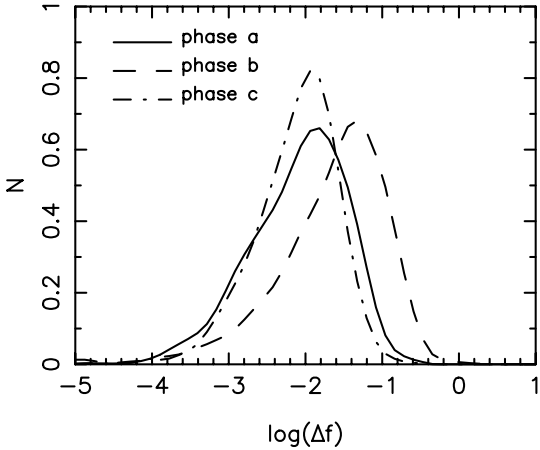}
\caption{Histograms of frequency drift parameter $\log(\Delta f)$ for
the three phases of model SA1 as indicated by the line-legends.
\label{fig:SA1_deltaf}}
\end{figure}

Figure~\ref{fig:SA1_deltaf} shows kernel density histograms of the
chaotic drift parameter $\log(\Delta f)$ for orbits in each of the
three phases in model SA1. It is obvious that in {\it phases a} and
{\it c} there is only a small fraction of chaotic orbits (i.e. orbits
with $\log(\Delta f) > -1$), whereas a much more significant fraction
of orbits lie to the right of this value in {\it phase b}. Even the
peak of the distribution in {\it phase b} is quite significantly
shifted to higher drift values.

In this section we investigate the surprising evidence that the
chaotic orbits in {\it phase b} of model SA1 do not appear to mix.
One possible reason for the lack of diffusion of the chaotic orbits is
that the timescale for evolution is not long enough. Indeed D08 report
that evolving run SA1 with the disk at full mass for an additional 5
Gyr after the growth of the disk is complete, leads to a larger
irreversible evolution (see their Figure 3a).  Nonetheless, even in
that case the irreversible evolution was only marginally larger than
when the disk was evaporated right after it grew to full mass.
Moreover the growth time was 5 Gyr which means that the halo was
exposed to a massive disk for a cosmologically long time.

A second possible reason for the lack of chaotic diffusion is that
most of the chaotic orbits in this phase of the simulation are
``sticky''.  The properties of ``sticky chaotic orbits" were discussed
in \S~\ref{sec:frequency}. In a series of experiments designed to
measure the rate of chaotic mixing, \citet{merritt_valluri_96} showed
that while ensembles of strongly chaotic orbits diffused and filled an
equipotential surface on timescales between 30-100 dynamical times,
similar ensembles of ``sticky" or resonantly trapped orbits diffused
much less quickly and only filled a small fraction of the allowed
surface after very long times.

Laskar (1990) showed that frequency maps are a powerful way to
identify resonances in dynamical systems.  Frequency maps are obtained
by plotting ratios of the 3 fundamental frequencies for each
individual orbit. If a large and representative orbit population is
selected, they can provide a map of the phase space structure of the
potential including all the resonances.  Resonances appear as straight
lines on the frequency map since their fundamental frequencies satisfy
a condition like $l\omega_x+m\omega_y+n\omega_z=0$. This method of
mapping the phase space has the advantage that since it only depends
on the ratios of the frequencies and not on the frequencies
themselves, it can be used to map phase space for large ensembles of
particles without requiring them to be iso-energetic. This is a
significant advantage over mapping schemes like Poincar\'e
surfaces-of-section, when applied to an $N$-body simulation where
particles, by design, are initialised to be smoothly distributed in
energy. Thus one can use the method to identify global resonances
spanning a large range of orbital energies in $N$-body simulations.

\begin{figure*}
\centering
\includegraphics[width=0.48\textwidth]{orbiSA1_mapE.ps}
\includegraphics[width=0.48\textwidth]{orbPlA3_mapE.ps}
\includegraphics[width=0.48\textwidth]{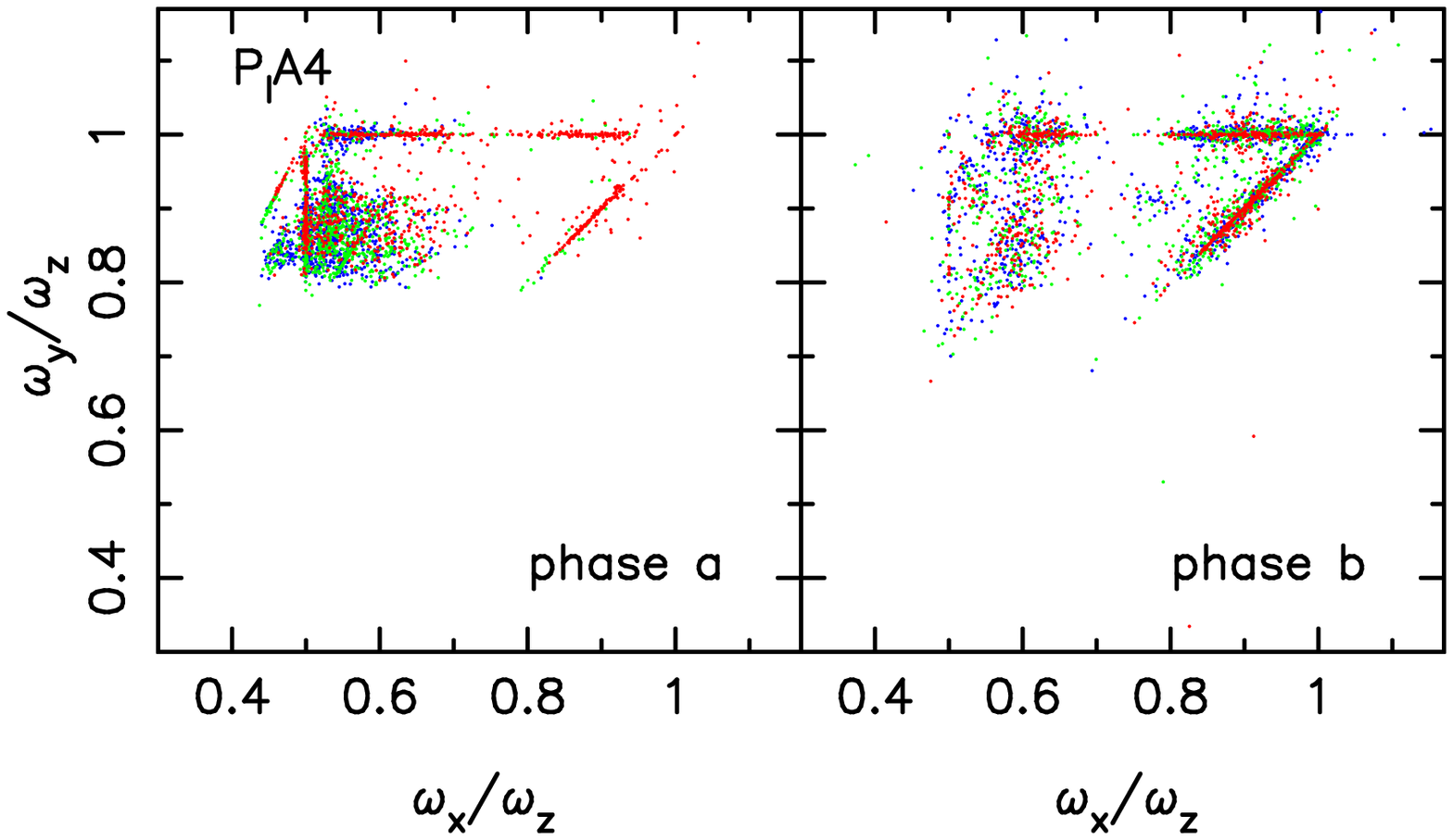}
\includegraphics[width=0.48\textwidth]{orbPlB3_mapE.ps}
\caption{Frequency maps of particles in {\it phase a} and {\it phase b}
for four models. For each particle the ratio of the
fundamental frequencies $\omega_y/\omega_z$ is plotted versus
$\omega_x/\omega_z$ is plotted by a single dot. The dots are colour coded by the energy  of the particle in {\it phase a}. The most tightly bound particles are coloured blue, and the least bound
particles are coloured red. Model SA1 has 6000 particles, model \PlA4 and \PlhA3 
have 5000 particles, while \PlB3 has 1000 particles each.
\label{fig:all_maps}}
\end{figure*}

In Figure~\ref{fig:all_maps} we present frequency maps for four of the
five models in {\it phase a} (left panels) and {\it phase b} (right
panels) (\PfB2 is not shown since the frequency maps for this model
are indistinguishable from those for \PlB3). For each orbit the ratios
of the fundamental frequencies $\omega_y/\omega_z$ and
$\omega_x/\omega_z$ are plotted against each other.  Particles are
colour coded by their energy in {\it phase a}. The energy range in
{\it phase a} was divided into three broad energy bins, with equal
numbers of particles per bin. The most tightly bound particles are
coloured blue, the least bound particles are coloured red and the
intermediate energy range is coloured green.

Resonance lines are seen in the clustering of particles in all the
maps. The most striking of the frequency maps is that for {\it phase
b} of model SA1 (\td). This map has significantly more prominent
resonance lines, around which many points cluster, than any of the
other maps.  Three strong resonances and several weak resonances are
clearly seen as prominent straight lines.
The horizontal line at $\omega_y/\omega_z = 1$ corresponds to the
family of orbits associated with the 1:1 closed (planar) orbit that
circulates around the $x$-axis, namely the family of ``thin shell''
L-tubes. The diagonal line running from the bottom left corner
to the top right corner with a slope of unity ($\omega_y/\omega_z =
\omega_x/\omega_z$) corresponds to the family of orbits that
circulates about the $z$-axis: the family engendered by the ``thin
shell'' S-tubes. Since this latter family shares the symmetry
axis of the disk, it is significantly strengthened in {\it phase b}
by the growth of the disk. In addition to having many more orbits
associated with it, this resonance extends over a much wider range in
energy as evidenced by the color segregation along the resonance line
(blue points to the bottom left and red points at the top right).
This segregation is the result of an increase in the gradient of the
potential along the $z$-axis due to the growth of the disk, which
results in an increase in $\omega_z$.  The more tightly bound a
particle, the greater the increase in $\omega_z$, and the greater the
decrease in both its ordinate and abscissa.  The most bound particles
(blue points) therefore move away from their original positions
towards the bottom left hand corner of the plot.  The least bound
particles (red points) are furthest from the center of the potential
and these points experience the least displacement - although these
points also shift slightly toward the resonance lines.  


\begin{figure*}
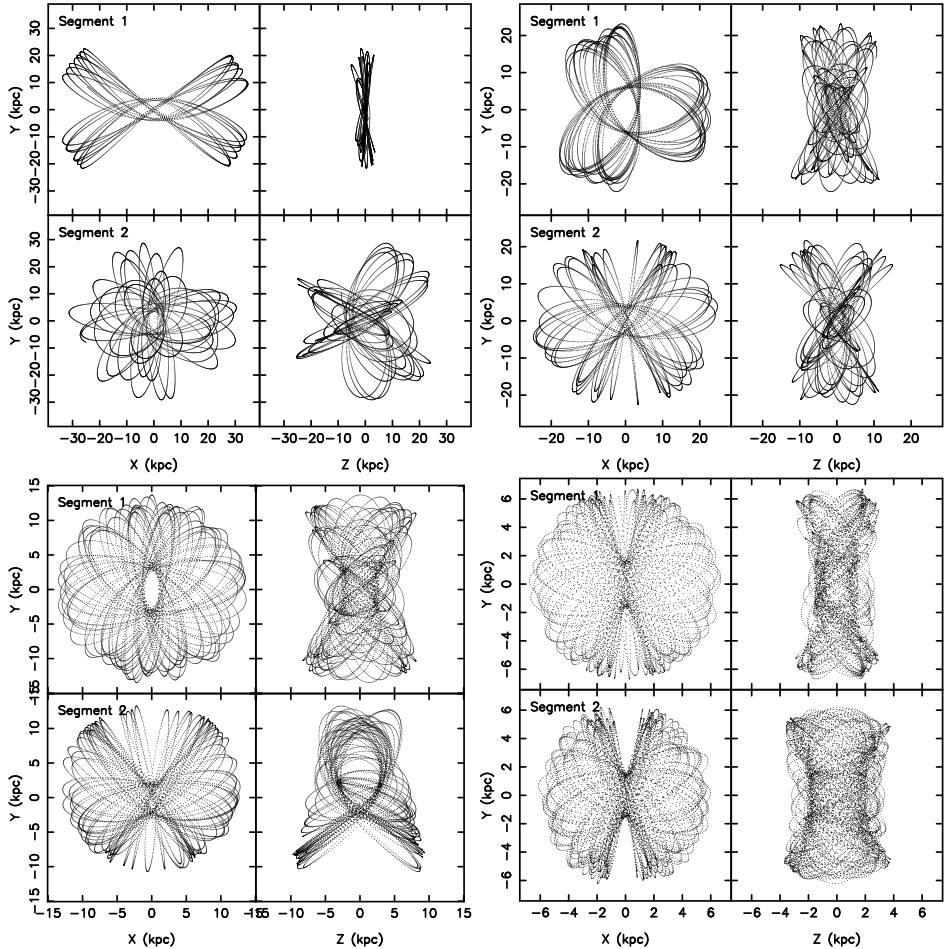
 
\centering
\includegraphics[width=0.35\textwidth]{orbiSA1_orb_x812.ps}
\includegraphics[width=0.35\textwidth]{orbiSA1_orb_x933.ps}
\includegraphics[width=0.35\textwidth]{orbiSA1_orb_x449.ps}
\includegraphics[width=0.35\textwidth]{orbiSA1_orb_x708.ps} 
\caption{Several chaotic orbits in
{\it phase b} of model SA1. Orbits are plotted in two Cartesian
projections over two different time segments of 10~Gyr (top two plots
in each panel show the first time segment, while bottom panels show
the second time segment).  See text for details.
\label{fig:chaotic-ex}} \end{figure*} 

A third prominent resonance is the vertical line at $\omega_x/\omega_z
= 0.5$ that corresponds to orbits associated with the family of banana
(1:2 resonant) orbits.  This banana (boxlet) resonance is also
enhanced by the growth of the disk since this family of orbits, while
not axisymmetric, is characterised by large excursions along the
$x$-axis and smaller excursions in the $z$ direction. Several shorter
resonance lines are seen but are too sparsely populated in this plot
to properly identify.

The frequency map for model \PlhA3 in {\it phase b} shows that a
spherical baryonic component produces a rather different phase space
structure than that produced by the disk.  In particular it is
striking that the most tightly bound (blue) points are now clustered
at the intersection of the horizontal and diagonal resonances namely
around the closed period orbits 1:1:1. This may be understood as the
consequence of the growth of the spherical baryonic point mass around
which all orbits are rosettes and since no direction is preferred all
orbits are ``round". The 1:2 banana resonance is also less prominent
in this model (largely because the deep central potential destabilises
this boxlet family).

The frequency map for model \PlA4 {\it phase b} shows the greatest
degree of scattering, as evidenced by the thickest resonance lines. We
attribute this to the large number of chaotic orbits in this
model. Apart from the broad clustering of points around the diagonal
(S-tube) and horizontal (L-tube) resonances there are no strong
resonance lines seen in this map. Unlike the map for \PlhA3 which
shows a clustering of tightly bound (blue) points at the 1:1:1
periodic orbit resonances, the blue points are widely scattered in the
frequency map of \PlA4.

The frequency maps for model \PlB3 shows that most of the orbits in
this model are associated with the (1:1) L-tube family (horizontal
line). A smaller number of orbits is associated with the 1:1 S-tube
resonance (diagonal line).  We saw previously that the growth of the
baryonic components in this prolate halo caused little change in the
orbit families. This is confirmed by the fact that the frequency maps
in both phases are remarkably similar except for an increase in the
clustering of points at the intersection of the horizontal (L-tube)
and the diagonal (S-tube) resonance, which occurs for the same reason
as in \PlhA3.  Since halo B is initially a highly prolate model, it
has (as we saw previously) only a small fraction of box (and boxlet)
orbits and in particular no banana orbits.

It is quite striking that in {\it phase a} the frequency maps show
significantly less segregation by energy, and only a few
resonances. This is because the initial triaxial models were generated
out of mergers of spherical NFW halos which were initially constructed
so that orbits were smoothly distributed in phase space. The increase
in the number of resonances following the growth of a baryonic
component is one of the anticipated consequences of resonant trapping
that occurs during the adiabatic change in a potential
\citep[e.g.][]{tre_yu_00, BT08}.

To test the conjecture that the majority of chaotic orbits in model
SA1 ({\it phase b}) are resonantly trapped, we compute the number of
chaotic orbits that lie close to a major resonance line. We define
``closeness'' to the resonance by identifying those orbits whose
frequency ratios lie $\pm \alpha$ of the resonant frequency ratio. For
example we consider an orbit to be close to the (1:1) L-tube resonance
(horizontal line in map), if $ |\omega_y/\omega_z-1| \le \alpha$. We
find that the fraction of chaotic orbits in {\it phase b}, that lie
close to one of the three major resonances identified above, is 51\%
when $\alpha = 0.01$ and 62\% when $\alpha = 0.03$.  Weaker resonances
lines (which are hard to recognise due to the sparseness of the data
points) may also trap some of the chaotic orbits. This supports our
conjecture that the main reason that model SA1 does not evolve in {\it
phase b}, despite the presence of a significant fraction of chaotic
orbits, is that the majority of the chaotic orbits are trapped around
resonances and therefore behave like regular orbits for very long
times.


In Figure~\ref{fig:chaotic-ex} we plot four examples of chaotic orbits
in {\it phase b} of SA1, which illustrate how resonantly trapped or
``sticky'' chaotic orbits look.  Each panel of four sub-plots shows a
single orbit plotted in two Cartesian projections (side-by-side). The
top pair of subplots show the orbit over the first 10 Gyr long time
segment, while the bottom pair shows the same orbit over a second
10~Gyr time segment. The two time segments were separated by
10~Gyr. For illustration we selected orbits with a range of drift
parameters.  The orbit in the top-left panel is an example of an orbit
that conforms to our notion of a chaotic orbit that explores more
phase space as time progresses, and has a large drift parameter of
$\log(\Delta f) = -0.48$. The top-right panel shows a S-tube orbit
that suddenly migrates to a box orbit (this orbit has a $\log(\Delta
f) = -0.54$) and was probably in the separatrix region between the
S-tube and box families.  The bottom-left panel shows an orbit that is
originally a S-tube that becomes trapped around a resonant boxlet
(``fish'') family (with $\log(\Delta f) = -0.66$), while the bottom
right-hand panel shows a weakly chaotic box orbit (with $\log(\Delta
f) = -0.94$). Of the 21\% of orbits in {\it phase b} that are chaotic
($\log(\Delta f) \ge -1$), only $\sim 5$\% have $\log(\Delta f) \ge
-0.5$. This fraction is small enough that one does not expect it to
result in significant chaotic mixing.


\section{Summary and Discussion }
\label{sec:summary}

Since it was first proposed, the idea that a central black hole would
scatter centrophilic box orbits in triaxial galaxies resulting in more
axisymmetric potentials \citep{ger_bin_85} has frequently been used to
explain the shape change in a variety of systems from the destruction
of bars by central black holes \citep{norman_etal_96} to the formation
of more oblate galaxy clusters in simulations with gas
\citep{kkzanm04}.

Experiments on chaotic mixing indicated that the timescales for such
mixing is about 30 - 100 dynamical times \citep{merritt_valluri_96},
which is much longer than the timescales for evolution of dark matter
halos in simulations with gas \citep{kkzanm04}. In addition recent
detailed studies of $N$-body simulations with controlled experiments
have shown that the role of chaotic mixing may be less dramatic than
conjectured by these previous studies.  A study of relaxation of
collisionless systems following the merger of two spherical galaxies
showed that despite the fact that a large fraction of the orbits in a
system undergoing violent relaxation are chaotic, the timescales for
chaotic diffusion and mixing are too long for this process to play a
significant role \citep{valluri_etal_07}. In fact, even after violent
relaxation, orbits retain strong memories of their initial energies
and angular momenta.

D08 argued that since chaos introduces an irreversible mixing,
numerical experiments in which evolution is driven by chaotic orbits
should not be reversible. These authors studied the macroscopic shapes
of triaxial dark matter halos in response to the growth of a baryonic
component. Unless the baryons were too centrally concentrated, or
transported angular momentum to the halo, the evolution they saw was
reversible, from which they concluded that much of the shape change
arises from deformations in the shape of individual orbits rather than
significant chaotic scattering.  In this paper we investigated this
issue in significantly greater detail by applying the Numerical
Analysis of Fundamental Frequencies (NAFF) technique that allows us to
quantify the degree to which chaotic diffusion drives evolution and to
identify the primary physical processes that cause halo shape change.
The frequency based method is able to distinguish between regular and chaotic orbits, making it more useful than Lyapunov exponents which are known to be sensitive to discretization effects in $N$-body systems \citep{hemsendorf_merritt_02}. We use the method to quantify the drift in frequencies of large
representative samples of orbits, thereby quantifying the degree of
chaos in the systems we study. It also allowed us to map the phase space
structure of the initial and final halos and to quantify the
relationship between the change in the shapes of individual orbits and
the shape of the halo as a whole.

Applying various analysis methods to orbits in five systems we
demonstrated that the conclusion reached by D08 that chaos is not an
important driver of shape evolution when the baryonic component is
extended is indeed valid. As did D08, we also found that significant
chaotic scattering does occur when the baryonic component is in the
form of a hard central point mass (of scale length $\sim 0.1$~kpc).
It is interesting that regardless of the original orbital composition
of the triaxial or prolate halo, and regardless of the shape of radial
scale length of the baryonic component, halos become more oblate
following the growth of a baryonic component.  Thus two quite
different processes (chaotic scattering and adiabatic deformation)
result in similar final halo shapes even in halos with very different
orbital compositions.

We explored two different initial halos, one in which box orbits were
the dominant elongated population (halo A) and the other in which
L-tubes dominated the initial halo (halo B). Despite the different
orbit compositions both models exhibit similar overall evolution with
regard to the shapes of orbits.  In the halo A models, the box orbits
were much more likely to change to either L-axis or S-tubes, whereas
in the halo B models, the dominant family of L-tubes largely retained
their orbital classification while deforming their shapes.

Below we list the main results of this paper:
\begin{enumerate}
\item Correlations between the orbital frequencies in the three
different phases $\omega_a$, $\omega_b$ and $\omega_c$ are a useful
way to search for regular versus chaotic evolution of orbits.  The
orbital frequencies in the three phases are found to be strongly
correlated with each other when the baryonic component is extended
(Fig.~\ref{fig:soft_omega_abc}), but show significant scattering when
the baryonic component is a compact point mass (of scale length about
0.1~kpc) (Fig.~\ref{fig:hard_omega_abc}). In the more extended
distributions, only a small fraction of the orbits experience
significant change in their original orbital frequencies when the
baryons are evaporated, while both the magnitude of scattering in
orbital frequency as well as the fraction of orbits experiencing
scattering, increases as the baryonic component becomes more compact
(Fig.~\ref{fig:all_omega_change_hist}).

\item In the three models with relatively extended baryonic
components, the change in orbital frequency between {\it phase c} and
{\it phase a} ($\Delta \omega_{ac}$) is not correlated with orbital
frequency (Fig.~\ref{fig:all_domega_omegaa}), pericenter distance or
orbital angular momentum. When the baryonic component is a hard point
mass, however, the frequency change is greater for orbits that are
deeper in the potential and therefore have both a higher initial
orbital frequency (Fig.~\ref{fig:all_domega_omegaa}) and smaller
pericentric radius (Fig.~\ref{fig:Pl3_jtot_rperi}).  Scattering in
frequency affects both the centrophilic box orbits as well as
centrophobic L-tubes.

\item The growth of a baryonic component in halo A (either disk or
softened point mass) causes box orbits with large pericentre radii to
be converted to S-tubes, L-tubes or become chaotic
(Fig.\ref{fig:typab_rp} top panel). While this change is almost
completely reversible in the case of the disk or a diffuse point mass,
it is less so when the baryonic component is a hard point mass. In
halo B, which is dominated by L-tubes, the growth of the baryonic
component causes almost no change in the orbital composition of the
halo, indicating that the L-tubes are not destroyed but deformed
(Fig.\ref{fig:typab_rp} bottom panel).  Even though \PlB3 (\pbbh) is a
model with significant orbit scattering by the hard central point
mass, the process appears to mainly convert elongated inner L-tube
orbits to somewhat rounder outer L-tubes. In model \PlA4 (\tbbh) the
box orbits are scattered onto S-tubes or chaotic orbits. The
significant amount of scattering seen for even centrophobic L-tube
orbits shows that the evolution is not due to direct scattering by
a central point mass as sometimes assumed. Two alternative 
possibilities  are more likely to account for the significant 
scattering in frequency. First the change in the symmetry and 
depth of the central potential is a perturbation to the potential that 
gives rise to an increase in the region of phase space occupied 
by resonances (Kandrup 1998 - private communication). As the 
resonances overlap there is an increase in the degree of chaotic 
behavior \citep{chirikov_79}. The second option is that the point mass attains
equipartition with the background mass distribution, resulting in
Brownian motion \citep{merritt_2005}. The Brownian motion can cause
the centre of the point mass to wander within a region of radius $\sim
0.1-1$~kpc which can result in a significant change in the maximum
gravitational force experienced by an orbit from one pericentre
encounter to the next.  This change in the maximum gravitational force
manifests as scattering of the orbit which is equally effective for
both box orbits and long-axis tubes.  Indeed a small wandering of the
central massive point is seen in the $N$-body simulations of \PlA4;
this motion is not included in our orbit calculations since all
particles are frozen in place when calculating orbits.  While it is beyond
the scope of this paper to explore this issue further, we caution that the 
motion of the point mass in our $N$-body simulations is likely to 
over-estimate the magnitude of the Brownian motion, since this 
depends on the mass resolution of the background particles. This 
suggests that in a real galaxy the evolution of the shape is much 
more likely to be driven by smooth adiabatic deformation of orbits than chaotic scattering.

\item In triaxial halos, the orbital shapes sharply peaked
distribution with the most elongated orbits ($\chi_s > 0.25$) are
either boxes or L-tubes.  In the prolate halos the second peak at
$\chi_s \sim 0$ contains a third of the orbits and is composed of
squat outer L-tubes and some box orbits.  The growth of a baryonic
component of any kind causes orbits of all types to become ``rounder",
especially at small pericenter radii.  This change in orbital shape
distribution with radius is the primary cause of the change of halo
shapes in response to the growth of a baryonic component. This is
consistent with the findings of D08 who also found that the orbits in
the models became quite round.

\item The growth of a disk causes a large fraction of halo orbits to
become resonantly trapped around major resonances. The three most
important resonances are those associated with the 1:1 tube (thin
shell) orbit that circulates about the short axis in the $x-y$-plane,
the 1:1 tube (thin shell) orbit that circulates about the long axis in
the $y-z$-plane and the 1:2 banana resonance in the $x-z$-plane. We
saw from the frequency maps that the resonant trapping of the halo
particles depends both on the form of the baryonic component grown in
the halo as well as on the initial orbital population of the halo.

\end{enumerate}

Thus we conclude that the evolution of galaxy and halo shapes
following the growth of a central component occurs primarily due to
regular adiabatic deformation of orbital shapes in response to the
changing central potential.  Chaotic scattering of orbits may be
important particularly for orbits with small pericentre radii but only
when the central point mass is extremely compact.  Contrary to
previous expectations, chaotic scattering is only slightly more
effective for centrophilic box orbits than it is for centrophobic
L-tubes. Boxes can be scattered onto both L- and S-tube orbits and a
significant fraction become chaotic. When the compact central point
mass scatters L-tubes as it does in a prolate halo, they are
scattered onto other L-tube orbits rather than onto S-tube orbits. The
strong chaotic scattering that we see on centrophobic L-tube orbits
has not been previously anticipated. 

An important implication of our analysis is that while the
shapes of halos (and by extension elliptical galaxies) become more
oblate (especially at small radii), following the growth of a baryonic
component, the majority of their orbits are not S-tubes as might be
predicted from their shapes. Instead our analysis shows that orbits
prefer to maintain their orbital characteristics, and the majority of
the orbits are those which would be generally found in triaxial
galaxies.  This is particularly important for studies of the internal
dynamics of elliptical galaxies since the fact that their shapes
appear nearly axisymmetric need not imply that their orbital structure
is as simple as the structure of oblate elliptical galaxies. Modifying
the shapes to slightly triaxial could result in significant changes in
their orbit populations and consequently could affect both the
inferred dynamical structure as well as the estimates of the masses
components such as the supermassive black holes in these galaxies
\citep{vandenbosch_dezeeuw_09}.

Finally, our finding that the growth of a stellar disk can result in a
large fraction of halo orbits becoming trapped in resonances could
have important implications for observational studies of the Milky
Way's stellar halo. The computation expense of the orbit calculations
forced us to restrict the size of the frequency map for model SA1 to
6000 particles. This is only a tiny fraction of the particles in the
original simulation. Despite the smallness of the sample, the
frequency maps (Fig.~\ref{fig:all_maps}) shows a rich resonant
structure which implies that the particles (either stars or dark
matter) in the stellar and dark matter halos of our Galaxy,
particularly those close to the plane of the disk, are likely to be
associated with resonances, rather than being smoothly distributed in
phase space (this is in addition to structures arising due to tidal
destruction of dwarf satellites). Although significantly greater
resolution is required to resolve such resonances than is currently
available, this could have significant implications for detection of
structures in current and upcoming surveys of the Milky Way such as
SDSS-III (Segue) and Gaia \citep{perryman_etal_01,wilkinson_etal_05}
and in on-going direct detection experiments which search for dark
matter candidates.

\section*{Acknowledgments}

M.V. is supported by the University of Michigan.  V.P.D. thanks the
University of Z\"urich for hospitality during part of this project.
Support for one of these visits by Short Visit Grant \# 2442 within
the framework of the ESF Research Networking Programme entitled
'Computational Astrophysics and Cosmology' is gratefully acknowledged.
Support for a visit by M.V. to the University of Central Lancashire at
an early stage of this project was made possible by a Livesey Grant
held by V.P.D.  All simulations in this paper were carried out at the
Arctic Region Supercomputing Center. We would like to thank the 
referee Fred Adams for his very thoughtful and constructive report 
which helped improve the paper.

\bibliographystyle{mn2e}
\bibliography{Master}

\bsp

\label{lastpage}

\end{document}